\documentclass[
reprint,
superscriptaddress,
amsmath,amssymb,
aps,
prb,
floatfix,
showkeys,
]{revtex4-2}

\usepackage{graphicx}% Include figure files
\usepackage{dcolumn}% Align table columns on decimal point
\usepackage{bm}% bold math
\usepackage[english]{babel}

\usepackage{slashed}
\usepackage{subcaption}
\usepackage[utf8]{inputenc}
\usepackage{textgreek}
\DeclareUnicodeCharacter{2212}{\textminus}
\usepackage[colorlinks=true,
            linkcolor=blue,
            urlcolor=blue,
            citecolor=blue]{hyperref} 
\begin{document}

%\preprint{APS/123-QED}

\title[Electronic structure of complex perovskite heterostructures at atomic-resolution EELS]{Quantitative near-atomic-resolution oxidation-state mapping in complex oxide heterostructures by electron energy-loss spectroscopy}
%{The Electronic Structure of \texorpdfstring{La$_{0.67}$Sr$_{0.33}$MnO$_3$/LaCr$_{0.2}$Mn$_{0.2}$Fe$_{0.2}$Co$_{0.2}$Ni$_{0.2}$O$_3$/SrTiO$_3$}{La$_{0.67}$Sr$_{0.33}$MnO3/LaCr$_{0.2}$Mn$_{0.2}$Fe$_{0.2}$Co$_{0.2}$Ni$_{0.2}$O3/SrTiO3} and \texorpdfstring{NdNiO$_3$/La$_{0.67}$Sr$_{0.33}$MnO$_3$/NdGaO$_3$}{La$_{0.67}$Sr$_{0.33}Mn$O3/NdNiO3} Perovskites Studied by Atomic-Resolution EELS}% Force line breaks with \\

\author{Ioannis Iatrakis}
 \email{ioannis.iatrakis@thermofisher.com}
\affiliation{Materials and Structural Analysis Division, Thermo Fisher Scientific,
Achtseweg Noord 5, 5651 GG, Eindhoven, The Netherlands}%

\author{Zitao Zhang}
 \email{zitao.zhang@thermofisher.com}
\affiliation{Materials and Structural Analysis Division, Thermo Fisher Scientific,
Achtseweg Noord 5, 5651 GG, Eindhoven, The Netherlands}%

\author{Ricardo Egoavil}
 \email{ricardo.egoavil@thermofisher.com}
\affiliation{Materials and Structural Analysis Division, Thermo Fisher Scientific,
Achtseweg Noord 5, 5651 GG, Eindhoven, The Netherlands}%

\author{Emma van der Minne}
\email{e.vanderminne@utwente.nl}
\affiliation{MESA+ Institute, Faculty of Science and Technology, University of Twente, 7500 AE Enschede, The Netherlands}%

\author{Christoph Baeumer}
\email{c.baeumer@utwente.nl}
\affiliation{MESA+ Institute, Faculty of Science and Technology, University of Twente, 7500 AE Enschede, The Netherlands}%

\author{Frank de Groot}
 \email{fmfdegroot@uu.nl}
\affiliation{Debye Institute for Nanomaterials Science, Utrecht University,  Universiteitsweg 99, 3584 CG, Utrecht, The Netherlands}%

\date{\today}
\begin{abstract} 
We present a theoretical framework for quantitative oxidation state mapping of transition-metal and rare-earth elements from near-atomic-resolution electron energy-loss spectroscopy (EELS) in complex oxide heterostructures. The total inelastic differential cross section of transition-metal L$_{2,3}$ and rare-earth M$_{4,5}$ edges is calculated by combining the relaxed Dirac-Hartree-Fock ionization cross section with ligand-field multiplet theory for the discrete transitions. We apply the approach to two correlated-oxide heterostructures, La$_{0.67}$Sr$_{0.33}$MnO$_3$/LaCr$_{0.2}$Mn$_{0.2}$Fe$_{0.2}$Co$_{0.2}$Ni$_{0.2}$O$_3$/SrTiO$_3$ and NdNiO$_3$/La$_{0.67}$Sr$_{0.33}$MnO$_3$/NdGaO$_3$, which include a wide range of transition-metal and rare-earth elements with overlapping features. The method resolves the oxidation states of Cr, Mn, Fe and Co and identifies the Ni contribution within the LaCr$_{0.2}$Mn$_{0.2}$Fe$_{0.2}$Co$_{0.2}$Ni$_{0.2}$O$_3$ layer at near-atomic resolution. We find predominantly Co$^{2+}$ together with mixed Mn$^{2+}$/Mn$^{3+}$/Mn$^{4+}$ character, indicating site-specific valence redistribution within the layer. We also separate overlapping spectral features, such as La-M$_{4,5}$ and Ni-L$_{2,3}$ edges in the NdNiO$_3$/La$_{0.67}$Sr$_{0.33}$MnO$_3$/NdGaO$_3$ heterostructure, thereby recovering the intrinsic Ni-L$_{2,3}$ excitation spectrum. We find that the effective Ni valence in NdNiO$_3$ deviates from the simple ionic picture. More broadly, this work establishes a generic method for quantitative valence mapping from near-atomic-resolution EELS, applicable to a wide range of experimental conditions for the study of local electronic structure of complex oxides.
\end{abstract}

%Use showkeys class option if keyword
                              %display desired}%Use showkeys class option if keyword
\maketitle
\newpage

%\tableofcontents

\section{\label{sec:Intro}Introduction}
Understanding the electronic structure of materials is essential for interpreting and tailoring their physical properties, including electrical and thermal conductivity, magnetic susceptibility, and catalytic activity. Transition-metal oxides are particularly important in this context because strong coupling between charge, spin, orbital, and lattice degrees of freedom gives rise to a broad range of emergent phenomena relevant to electronics, energy technologies, and catalysis.\cite{Tokura2000,Orenstein2000, baeumer2017anomalous, Catalano_2018, D3CC00325F, Baeumer2021}

A key parameter governing the behavior of transition-metal oxides is the oxidation state of the metal cation. In layered and interfacial systems, deviations from bulk valence states can arise from charge transfer, polar discontinuities, epitaxial strain, or oxygen nonstoichiometry. Well-known examples include the CuO$_2$ planes of cuprate superconductors, where the Cu oxidation state controls carrier concentration and superconducting properties,\cite{Orenstein2000} as well as oxide heterostructures in which interfacial charge transfer drives emergent electronic and magnetic behavior \cite{PhysRevLett.119.156801, https://doi.org/10.1002/admi.201801428, GarciaBarriocanal2010, doi:10.1021/acsnano.2c08096, kxnn-w1rk, Geessinck2020}.

La$_{0.67}$Sr$_{0.33}$MnO$_3$/LaCr$_{0.2}$Mn$_{0.2}$Fe$_{0.2}$Co$_{0.2}$Ni$_{0.2}$O$_3$/SrTiO$_3$ and NdNiO$_3$/La$_{0.67}$Sr$_{0.33}$MnO$_3$/NdGaO$_3$ heterostructures provide two complementary systems for interfacial valence analysis. In stoichiometric LaMnO$_3$, Mn is nominally high-spin Mn$^{3+}$ with a $t_{2g}^{3}e_g^{1}$ configuration, making the manganite strongly Jahn-Teller active. Octahedral distortions and orbital order are closely tied to its antiferromagnetic insulating ground state. Substitution of divalent Sr$^{2+}$ for trivalent La$^{3+}$ introduces holes into the Mn sublattice and increases the nominal average Mn oxidation state from $3+$ to $(3+x)+$ in La$_{1-x}$Sr$_x$MnO$_3$. For $x=0.33$, the ideal ionic picture therefore corresponds to a mixed-valence state containing approximately two-thirds Mn$^{3+}$ and one-third Mn$^{4+}$. This mixed valence reduces the average $e_g$ occupancy and enables ferromagnetic double-exchange interactions mediated by itinerant $e_g$ electrons. Consequently, La$_{0.67}$Sr$_{0.33}$MnO$_3$ exhibits a ferromagnetic metallic ground state over a broad temperature range. Because the magnetic and transport properties are highly sensitive to the Mn oxidation state, additional interfacial charge transfer, oxygen nonstoichiometry or mixing of Mn$^{2+}$, Mn$^{3+}$ and Mn$^{4+}$ states can substantially alter the local electronic state.  The Mn-L$_{2,3}$ edge is therefore a particularly sensitive probe of spatial variations in Mn oxidation state and electronic structure across these heterostructures.

The LaCr$_{0.2}$Mn$_{0.2}$Fe$_{0.2}$Co$_{0.2}$Ni$_{0.2}$O$_3$ high-entropy oxide (HEO) layer provides a chemically disordered multi-cation environment with multiple accessible oxidation states, enabling enhanced charge redistribution and local redox flexibility at the interface. For example, Co substitution on the Mn site in LaMn$_{1-x}$Co$_x$O$_3$ was reported to induce Mn$^{3+}$/Mn$^{4+}$ mixed valence, with the resulting ferromagnetism interpreted in terms of Mn$^{3+}$–Mn$^{4+}$ double exchange \cite{PhysRevB.55.11072}.

On the SrTiO$_3$ side, the Ti-L$_{2,3}$ and O-K fine structures provide a nominally Ti$^{4+}$, $d^0$ spectroscopic reference, whereas previous studies of La$_{1-x}$Sr$_x$MnO$_3$/SrTiO$_3$ have shown that Ti can remain predominantly Ti$^{4+}$ while the Mn valence on the La$_{1-x}$Sr$_{x}$MnO$_3$ side varies with thickness and charge redistribution \cite{10.1063/1.2202118, 736f0e9ea5ec46a6b536e4ef5bc98ae3, Trappen2018Electrostatic}.

NdNiO$_3$, by contrast, is a correlated nickelate whose electronic structure has been investigated in a full-correlation picture with intercluster coupling \cite{PhysRevB.94.195127}. The system is characterized by negative charge transfer and strong Ni-O covalency. The effective Ni valence is closer to ${2+}$, with charge being compensated by ligand holes, rather than a pure ${3+}$ state. The interface of NdNiO$_3$/La$_{0.7}$Sr$_{0.3}$MnO$_3$ has been investigated using x-ray absorption and x-ray magnetic circular dichroism spectra. Ni reduction at the interface  has been attributed to charge transfer involving Mn \cite{Chen2020}. The metallic and insulating phases of perovskite nickelates have also been studied by scanning transmission electron microscopy–electron energy-loss spectroscopy (STEM-EELS) in terms of the O-K prepeak width and empirical characterization of Ni-L$_{2,3}$ \cite{doi:10.1021/acs.nanolett.0c04538}.  The  quantitative analysis of  Ni spectral features in La-containing samples is a particularly demanding task, because the most informative part of the Ni-L$_3$ edge overlaps strongly with La-M$_4$, so quantitative analysis requires explicit La/Ni deconvolution \cite{PhysRevB.90.195140}.

Electron energy-loss spectroscopy in the scanning transmission electron microscope is an element-specific probe of excited electronic states and is highly sensitive to oxidation state, crystal-field splitting, and local symmetry through the near-edge fine structure of core-loss edges \cite{PhysRevB.32.5107, Abbate1992, deGroot2005, degroot2008, Kourkoutis14122010, doi:10.1021/acs.inorgchem.2c02830, GLOTER20172}. Aberration-corrected STEM-EELS has further enabled the spatial mapping of composition and electronic structure with atomic spatial resolution,\cite{BROWNING1997333, 1999Natur.399..758M, doi:10.1126/science.1148820, Krivanek_2010, PhysRevLett.107.107602, KRIVANEK201960, refId0, chien2013visualizing}. The quantitative interpretation of transition-metal L$_{2,3}$ and rare-earth M$_{4,5}$ edges is strongly influenced by the core hole in the final state and by multiplet effects arising from the many-electron character of the core-excited states. These edges are excitonic in nature and, in general, cannot be interpreted directly in terms of the unoccupied density of states. As a result, ground-state methods alone do not fully capture the relevant spectral complexity, while empirical fingerprinting approaches can be sensitive to the availability and quality of appropriate experimental reference spectra, especially in systems exhibiting mixed valence and severe edge overlap \cite{deGroot2005, degroot2008}.

In this work, we present a theoretical framework for quantitative oxidation-state analysis of transition-metal and rare-earth elements in near-atomic-resolution STEM-EELS. We analyze monochromated spectrum-image datasets from La$_{0.67}$Sr$_{0.33}$MnO$_3$/LaCr$_{0.2}$Mn$_{0.2}$Fe$_{0.2}$Co$_{0.2}$Ni$_{0.2}$O$_3$/SrTiO$_3$ and NdNiO$_3$/La$_{0.67}$Sr$_{0.33}$MnO$_3$/NdGaO$_3$ heterostructures using total inelastic differential cross sections calculated within the relaxed Dirac-Hartree-Fock framework, explicitly including core-hole effects, together with crystal-field and charge-transfer multiplet calculations \cite{IatrakisBrudanin2026}. Implemented in an automated pixel-by-pixel analysis, this approach resolves Mn$^{2+}$, Mn$^{3+}$, and Mn$^{4+}$ contributions in the La$_{0.67}$Sr$_{0.33}$MnO$_{3}$ and HEO layers, identifies Cr-, Fe-, and Co-related oxidation states present locally in the film, and deconvolves the strongly overlapping La-M$_{4,5}$ and Ni-L$_{2,3}$ edges in NdNiO$_3$/La$_{0.67}$Sr$_{0.33}$MnO$_{3}$ without empirical subtraction. The methodology enables robust oxidation-state mapping across oxide layers and interfaces and provides direct insight into local redox processes and electronic structure in complex oxides.

\section{Experimental methods}
La$_{0.67}$Sr$_{0.33}$MnO$_{3}$/LaCr$_{0.2}$Mn$_{0.2}$Fe$_{0.2}$Co$_{0.2}$Ni$_{0.2}$O$_{3}$ heterostructures were grown by RHEED–assisted PLD on B-site-terminated, step-terraced 0.5 wt\% Nb-doped SrTiO$_3$ (001) substrates (CrysTec GmbH, Germany). NdNiO$_3$ /La$_{0.67}$Sr$_{0.33}$MnO$_3$ were grown by RHEED–assisted PLD on B-site-terminated, step-terraced NdGaO$_3$ (110) substrates (CrysTec GmbH, Germany). The stoichiometric NdNiO$_3$ (SurfaceNet) and La$_{0.67}$Sr$_{0.33}$MnO$_3$ (SurfaceNet) targets were ablated using a KrF excimer laser (248 nm, Coherent). The LaCr$_{0.2}$Mn$_{0.2}$Fe$_{0.2}$Co$_{0.2}$Ni$_{0.2}$O$_{3}$  PLD target was synthesized via sintering of perovskite powder using reverse co-precipitation followed by a calcination process. The powders were annealed at 1200 °C for 7 h, followed by uniaxial pressing at 300 MPa and a heat treatment at 1500 °C for 12 h \cite{https://doi.org/10.1002/adma.202102301}. 
La$_{0.67}$Sr$_{0.33}$MnO$_{3-\delta}$ layers were deposited at 750$^\circ$C in 0.266 mbar O$_2$, LaCr$_{0.2}$Mn$_{0.2}$Fe$_{0.2}$Co$_{0.2}$Ni$_{0.2}$O$_{3}$ layers were deposited at 650$^\circ$C in 0.04 mbar O$_2$, while NdNiO$_3$ layers were grown at 650$^\circ$C in 0.2 mbar O$_2$. In both cases, a laser fluence of 2.0 J cm$^{-2}$ and a repetition rate of 1 Hz were used. After deposition, samples were cooled to room temperature at 25$^\circ$C min$^{-1}$ in 0.2 mbar O$_2$.\\\\
All depositions were carried out with a fixed target-to-substrate distance of 5 cm and a rectangular laser spot size of 2.24 mm$^2$, defined by a mask. Targets were pre-ablated at 10 Hz. Growth was performed in a TSST vacuum system (base pressure $5 \times 10^{-8}$ mbar) equipped with \textit{in situ} RHEED (Staib Instruments, Germany).\\

Near-atomic-resolution STEM-EELS spectrum imaging was performed using a Thermo Scientific ILIAD transmission electron microscope operated at 300~kV. The instrument is equipped with probe and image aberration correctors, a monochromated electron source, and an ILIAD energy filter with a Zebra detector comprising five detector arrays for multi-EELS acquisition.

The monochromator was configured to achieve an energy resolution of $\approx$ 400~meV, as determined from the full width at half maximum (FWHM) of the zero-loss peak, recorded with an energy dispersion of 169~meV per channel. A probe convergence semi-angle of 18~mrad was used, providing atomic resolving power and a beam current of $\approx$ 120~pA. The EELS collection semi-angle was set to 45~mrad.

STEM-EELS spectrum images were acquired for both La$_{0.67}$Sr$_{0.33}$MnO$_3$/LaCr$_{0.2}$Mn$_{0.2}$Fe$_{0.2}$Co$_{0.2}$Ni$_{0.2}$O$_3$/SrTiO$_3$ and NdNiO$_3$/La$_{0.67}$Sr$_{0.33}$MnO$_3$/NdGaO$_3$ heterostructures along the [001] zone axis, with real-space pixel sizes of 42.36~pm and $84.72$~pm, respectively, enabling atomic-scale spatial resolution across the oxide heterostructure. The relative thickness of both systems was estimated as $\frac{t}{\lambda} \approx 0.4$ with small variation across layers, using the log-ratio method, $\frac{t}{\lambda} = \ln(I_{tot} / I_{el})$, where $I_{tot}$ is the integrated elastic and inelastic signal and $I_{el}$ is the elastic signal.

To simultaneously probe multiple elemental edges with high energy resolution and optimized signal-to-noise ratio, the Multi-EELS acquisition mode of the Thermo Scientific ILIAD platform was employed. This acquisition mode allows recording different energy ranges sequentially within a single scan, with independently optimized exposure conditions. The Multi-EELS settings for La$_{0.67}$Sr$_{0.33}$MnO$_3$/LaCr$_{0.2}$Mn$_{0.2}$Fe$_{0.2}$Co$_{0.2}$Ni$_{0.2}$O$_3$/SrTiO$_3$ were as follows: an energy dispersion of 169~meV per channel was used for all energy ranges. The first energy range, starting at $-22$~eV, was acquired with an exposure time of 50~$\mu$s to capture the low-loss region. The second range, starting at 400~eV with an exposure time of 5~ms, was used to record the Ti, Cr, Mn, Fe L$_{2,3}$ and O-K edges. The third range, starting at 746~eV and acquired with a 5~ms exposure time, targeted the Co, Ni $L_{2,3}$ and La-M$_{4,5}$ edges. 
For the NdNiO$_3$/La$_{0.67}$Sr$_{0.33}$MnO$_3$/NdGaO$_3$ sample, an analogous acquisition scheme was used, with spectrum start energies $-22$~eV, 480~eV, 750~eV and exposure times 50~$\mu$s, 15~ms and 15~ms respectively. The low loss contains the zero loss peak and the M$_{2,3}$ and N$_{4,5}$ edges of the transition metal and rare earth elements. The second energy range covers the O-K and Mn-L$_{2,3}$ edges. The third energy range covers Ni-L$_{2,3}$ and the rare earth elements' edges La-M$_{4,5}$ and Nd-M$_{4,5}$.

This experimental configuration, combining monochromated illumination, near-atomic-resolution STEM Multi-EELS acquisition using the ILIAD energy filter, enables quantitative analysis of subtle variations in fine structure associated with mixed oxidation states at oxide interfaces.

\section{\label{sec:theory} Theoretical framework}

\subsection{Inelastic electron scattering cross section}

A quantitative interpretation of core-loss EELS requires a rigorous description of the inelastic interaction between the fast incident electron and the electronic system of the specimen. In this work, the inelastic scattering cross section is described within quantum electrodynamics following the formulation of \cite{IatrakisBrudanin2026}. The many-body effects of the core level ionization are incorporated in this framework.

\subsection{Ionization cross section}

We follow the generic framework for the inelastic scattering of relativistic electrons by atoms where final-state effects are explicitly included \cite{IatrakisBrudanin2026}. The interaction operator is expressed in terms of the electromagnetic four-potential, leading to a separation into
longitudinal (Coulomb) and transverse (retarded) contributions, where all relativistic retardation and spin-dependent terms are explicitly included.
This formulation naturally incorporates relativistic effects arising from the high velocity of the probe electron and becomes essential for quantitative description of core-loss excitations at large momentum transfer. The long wavelength limit of the cross section is applicable for low collection angle experiments and it is also developed in \cite{IatrakisBrudanin2026}. The atomic structure has been calculated using the relaxed Dirac-Hartree-Fock method with explicit calculation of the exchange interaction without requiring an effective central-field. Final-state effects are explicitly included in this formalism. We calculate the ionization part of the cross section in the relaxed Dirac-Hartree-Fock approach for all the ionization edges which are analyzed in the current work.

\subsection{Ligand-field multiplet theory}

The fine structure of transition-metal $L_{2,3}$ and rare-earth $M_{4,5}$ edges  is determined by the many-body electronic structure of the core-excited states. This structure is accurately described using ligand-field multiplet theory for ionic systems, while in more covalent systems charge transfer also contributes \cite{deGroot2005}.

In ligand-field multiplet theory, the final state created by a $2p \rightarrow 3d$ excitation is treated as a manifold of multiplet-split configurations arising from the strong Coulomb and exchange interactions between the $2p$ core hole and the partially filled $3d$ shell. The Hamiltonian explicitly includes intra-atomic electron-electron interactions, spin-orbit coupling in both the $2p$ and $3d$ shells, and crystal-field interactions that reflect the local symmetry and ligand coordination. For systems with small or negative charge-transfer parameter, charge-transfer configurations and metal-ligand hybridization must be included. This is important for high-valence systems such as Cu$^{3+}$ and Fe$^{4+}$ \cite{deGroot2005}. In some oxides the 3d spin-orbit coupling is important for the ground state (for example CoO), while in other oxides the 3d spin-orbit coupling is effectively quenched. When the 3d spin-orbit coupling is not quenched the appropriate thermal distribution over low-lying states at the measurement temperature should be included.

For octahedrally coordinated transition metals, the crystal field splitting parameter $10Dq$ plays a central role in determining the relative energies and intensities of multiplet features. Variations in oxidation state lead to systematic changes in multiplet structure, branching ratios, and edge onset energies, making the $L_{2,3}$ edges highly sensitive probes of valence and spin state \cite{Abbate1992}. To first order, moderate distortions from octahedral symmetry can often be neglected for oxidation-state identification, although the correct coordination symmetry must be used for larger changes such as tetrahedral coordination. For Mn$^{3+}$, Co$^{3+}$, and Ni$^{3+}$ in octahedral oxygen coordination, the spin state can be sensitive to the crystal-field strength; when it is not known independently, both high- and low-spin configurations should be tested. The coupling of the discrete spectrum with the continuum is also considered as it plays an important role for the high-energy tails of the rare-earth elements' spectral peaks \cite{PhysRevB.85.165113}.

In the current work, the excitation calculations are performed using Quanty, a software library for many-body theory calculations of transition-metal and rare-earth element spectra \cite{lu2014efficient, haverkort2014bands, PhysRevB.85.165113}. The coupling constants of the interaction Hamiltonian of the different oxidation states for each element are determined by theory and previous X-ray absorption spectroscopy (XAS) studies \cite{PhysRevB.42.5459, PhysRevB.85.165113, PhysRevB.41.928, PhysRevB.32.5107, MIEDEMA201332, d856bdb5612e449b98ef14b4c1b091d7}. For the collection geometry used, the excitation spectra are calculated within the dipole approximation since the dipole contribution dominates the excitation spectrum of transition metals and lanthanides \cite{d856bdb5612e449b98ef14b4c1b091d7}. The spectrum  is broadened by a Lorentzian kernel due to the natural lifetime of the excited states \cite{doi:10.1021/acs.inorgchem.2c02830}, and then convolved with the low-loss EELS spectrum to account for plural scattering. The relative energy of the ionization cross section with respect to the excitation spectrum has been determined using experimental results from X-ray photoelectron spectroscopy (XPS), XAS and EELS data \cite{PhysRevB.33.4253, doi:10.1021/acs.inorgchem.2c02830}. The discrete transition amplitudes are appropriately normalized following the angular momentum sum rules \cite{PhysRevLett.68.1943, DEGROOT1994529}. %Consequently, the relative fitted amplitudes of the different oxidation-state components directly yield their relative populations.

\section{Results}

\subsection{\texorpdfstring{La$_{0.67}$Sr$_{0.33}$MnO$_3$/LaCr$_{0.2}$Mn$_{0.2}$Fe$_{0.2}$Co$_{0.2}$Ni$_{0.2}$O$_3$/SrTiO$_3$}{La0.67Sr0.33MnO3/LaCr0.2Mn0.2Fe0.2Co0.2Ni0.2O3/SrTiO3}}

\begin{figure}[!htbp] 
%\begin{subfigure}[t]{.8\linewidth}
%\caption{}
\includegraphics[width=.6\linewidth]{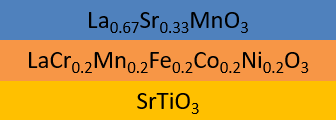}
%\end{subfigure}\hfill%
%\begin{subfigure}[t]{.8\linewidth}
%\caption{}
%\includegraphics[width=\linewidth]{images/LMOSTO/spectra/La-M_Area4_notitle.png}
%\end{subfigure}%
\caption{A schematic view of the multilayer sample.}
\label{fig:Lalmostoschematic}
\end{figure}

\begin{widetext}

\begin{figure}[!htbp]
    \centering
    \begin{subfigure}[b]{0.09\textwidth}
        \includegraphics[height=8cm, keepaspectratio]{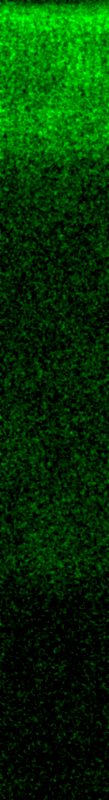}
        \caption{}
        \label{fig:mn2plus}
    \end{subfigure}
    \begin{subfigure}[b]{0.09\textwidth}
        \includegraphics[height=8cm, keepaspectratio]{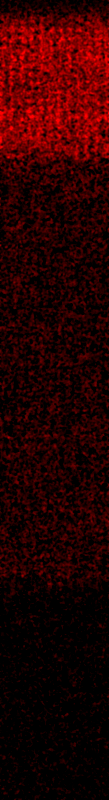}
        \caption{}
        \label{fig:mn3plus}
    \end{subfigure}
    \begin{subfigure}[b]{0.09\textwidth}
        \includegraphics[height=8cm, keepaspectratio]{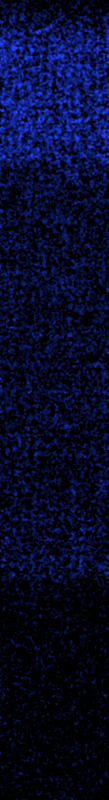}
        \caption{}
        \label{fig:mn4plus}
    \end{subfigure}
    \begin{subfigure}[b]{0.09\textwidth}
        \includegraphics[height=8cm, keepaspectratio]{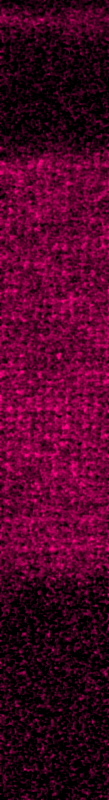}
        \caption{}
        \label{fig:cr3plus}
    \end{subfigure}
    \begin{subfigure}[b]{0.09\textwidth}
        \includegraphics[height=8cm, keepaspectratio]{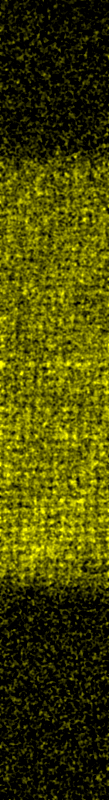}
        \caption{}
        \label{fig:fe3plus}
    \end{subfigure}
    \begin{subfigure}[b]{0.09\textwidth}
        \includegraphics[height=8cm, keepaspectratio]{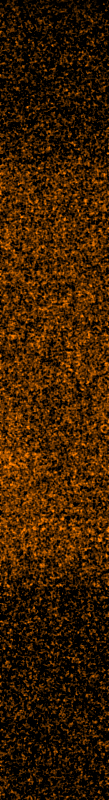}
        \caption{}
        \label{fig:co2plus}
    \end{subfigure}
    \begin{subfigure}[b]{0.09\textwidth}
        \includegraphics[height=8cm, keepaspectratio]{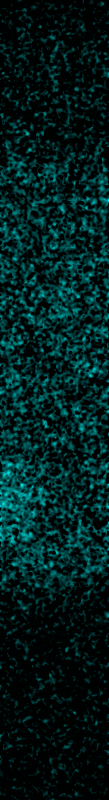}
        \caption{}
        \label{fig:ni2plus}
    \end{subfigure}
    \begin{subfigure}[b]{0.09\textwidth}
        \includegraphics[height=8cm, keepaspectratio]{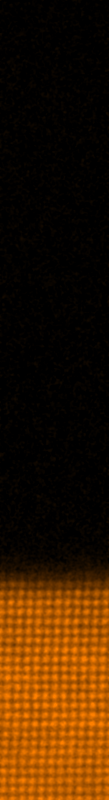}
        \caption{}
        \label{fig:ti4plus}
    \end{subfigure}
    \begin{subfigure}[b]{0.09\textwidth}
        \includegraphics[height=8cm, keepaspectratio]{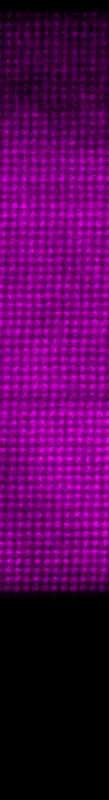}
        \caption{}
        \label{fig:la3plusmaplmo}
    \end{subfigure}
    \begin{subfigure}[b]{0.09\textwidth}
        \includegraphics[height=8cm, keepaspectratio]{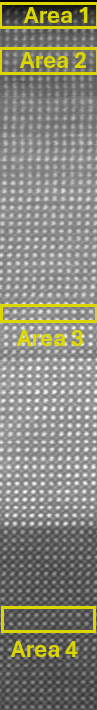}
        \caption{}
        \label{fig:dfslmo}
    \end{subfigure}
    \caption{The distribution maps of the different oxidation states on the spectrum image. (a) Mn$^{2+}$, (b) Mn$^{3+}$, (c) Mn$^{4+}$, (d) Cr$^{3+}$, (e) Fe$^{3+}$, (f) Co$^{2+}$, (g) Ni$^{2+, 3+}$ (h) Ti$^{4+}$, (i) La$^{3+}$. (j) contains the acquired dark-field STEM image. The three layers of La$_{0.67}$Sr$_{0.33}$MnO$_3$/LaCr$_{0.2}$Mn$_{0.2}$Fe$_{0.2}$Co$_{0.2}$Ni$_{0.2}$O$_3$/SrTiO$_3$ are located at the top (area 1 and 2), middle (area 3) and bottom (area 4) of the spectrum image respectively.}
    \label{fig:Mapslmo}
\end{figure}

\end{widetext}

We employ the total inelastic scattering cross section to extract oxidation-state maps from the spectrum-image data of the three layers of the sample shown in Figure \ref{fig:Lalmostoschematic} using Velox software. The spectral shapes of the individual oxidation states are fixed as described in Section \ref{sec:theory}, while their amplitudes are determined by fitting the experimental data. Since the theoretical cross sections are consistently normalized, the fitted amplitudes directly yield the relative populations of the different oxidation states of each element. The energy axis of each pixel spectrum is calibrated such that the zero-loss peak is centered at zero energy. Residual energy non-linearity in the core-loss range, due to aberrations,  is corrected by automated optimization of the spectrum dispersion during fitting. The theoretical spectral shapes are convolved with the low-loss spectrum to account for plural scattering and experimental broadening. The edge models and the EELS background are fitted simultaneously using weighted multivariate regression, with the weights determined from the Poisson counting statistics and readout noise of the experimental data. This integrated framework enables quantitative fitting of the experimental spectra and decomposition of mixed oxidation states at the pixel-spectrum level, resulting in the oxidation-state distribution maps shown in Figure \ref{fig:Mapslmo}. Prior to fitting, the spectrum-image data are spatially filtered with a Gaussian kernel of $\sigma=0.7$ pixels to improve the signal-to-noise ratio.

Because the Mn oxidation state directly controls the $e_g$ occupancy, even modest fractions of Mn$^{2+}$ or Mn$^{4+}$ can strongly modify the balance between Jahn-Teller distortion, superexchange, double exchange, and charge transport. The Mn valence is therefore a central control parameter for the local electronic and magnetic state of Mn-based perovskite films and heterostructures. Traditionally, determining transition-metal oxidation states from L$_{2,3}$-edge spectra has been challenging and has relied largely on empirical methods such as intensity-ratio measurements or comparisons with experimentally acquired spectral line shapes.  In contrast, our quantitative fitting based on theoretically calculated spectra for individual oxidation states enables distinct valence contributions to be identified and quantified, providing a more rigorous and sensitive analysis of local oxidation states than conventional qualitative evaluation. The relevant experimental parameters are incorporated directly into the theoretical calculation allowing the method to be applied in a wide range of experimental conditions.

\begin{figure}[!htbp] 
\begin{subfigure}[t]{.9\linewidth}
\caption{}
\includegraphics[width=\linewidth]{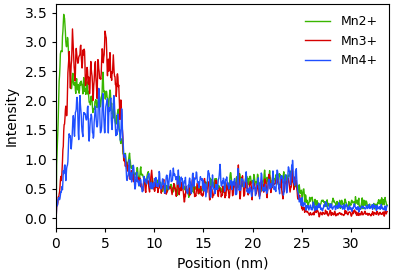}
\label{fig:lmoprofilemn}
\end{subfigure}\hfill%
\begin{subfigure}[t]{.9\linewidth}
\caption{}
\includegraphics[width=\linewidth]{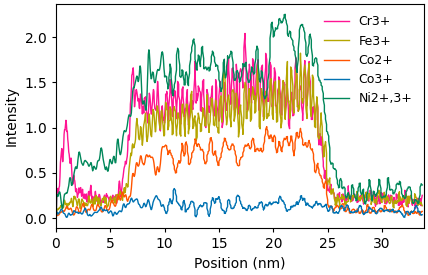}
\label{fig:lmoprofileheo}
\end{subfigure}%
\caption{(a) Intensity profiles of the fitted Mn$^{2+, 3+, 4+}$ oxidation state models along the vertical direction of the spectrum image shown in Fig.~\ref{fig:Mapslmo}. (b) Intensity profiles of the fitted Cr$^{3+}$, Fe$^{3+}$, Co$^{2+,3+}$, Ni$^{2+,3+}$ components along the vertical direction of the spectrum image. Ni$^{2+,3+}$ denotes the total Ni signal from both the 2+ and 3+ components.}
\label{fig:lmoprofiles}
\end{figure}

Figs. \ref{fig:mn2plus}, \ref{fig:mn3plus}, \ref{fig:mn4plus} and \ref{fig:lmoprofiles} reveal a pronounced vertical redox gradient across the La$_{0.67}$Sr$_{0.33}$MnO$_3$/LaCr$_{0.2}$Mn$_{0.2}$Fe$_{0.2}$Co$_{0.2}$Ni$_{0.2}$O$_3$/SrTiO$_3$ heterostructure. The fitted oxidation-state maps show that the La$_{0.67}$Sr$_{0.33}$MnO$_3$ and the high-entropy oxide layer contain spatially varying Mn$^{2+}$, Mn$^{3+}$, and Mn$^{4+}$ components. Quantitative oxidation-state mapping at atomic-scale resolution is achieved by fitting the theoretically calculated spectral components to each pixel of the EELS spectrum image, with the fitting procedure explicitly accounting for the Poisson counting statistics of the measured spectra.

The Mn$^{2+}$ contribution is strongest towards the top of the La$_{0.67}$Sr$_{0.33}$MnO$_3$ layer, consistent with a more reduced surface region, while Mn$^{3+}$ and Mn$^{4+}$ dominate the interior of the La$_{0.67}$Sr$_{0.33}$MnO$_3$ and HEO layers \cite{doi:10.1021/acsami.3c17639}. Figure \ref{fig:mnspeclmosto} shows the fit of the theoretical Mn model to the experimental spectra extracted from areas 1, 2, and 3 of the spectrum image, as shown in Fig.~\ref{fig:dfslmo}. The background of the EELS spectra is subtracted from all the spectrum plots to emphasize the characteristic spectral features. Figs.~\ref{fig:Mnspecarea1}, \ref{fig:Mnspecarea2}, and \ref{fig:Mnspecarea3} show the inelastic cross section together with the background-subtracted Mn spectrum from areas 1, 2, and 3, respectively. The ligand-field multiplet calculations of the Mn$^{2+,3+,4+}$ excitation spectra with $O_h$ symmetry reproduce the experimental data quite well. Mn$^{3+}$ is treated in the high-spin state, while Mn$^{4+}$ is described by the octahedral $3d^3$ ($t_{2g}^3$) configuration including charge transfer. Charge-transfer effects and the impact of distortions on the Mn$^{3+}$ and Mn$^{4+}$ spectra are discussed in detail in \cite{doi:10.1021/acs.inorgchem.2c02830}. In Figs.~\ref{fig:Mnexcarea1}, \ref{fig:Mnexcarea2}, and \ref{fig:Mnexcarea3}, the ionization part of the spectrum is also removed and the excitation spectrum is shown together with the individual Mn$^{2+,3+,4+}$ spectral components. The fitted amplitudes determine the relative populations of the different Mn oxidation states within each integration area. In area 1, the fitted Mn$^{2+}$, Mn$^{3+}$, and Mn$^{4+}$ populations are 58\%, 30\%, and 12\%, respectively, showing a clear reduction in the average Mn oxidation state. In area 2, the corresponding populations are 32\%, 40\%, and 28\%, while in area 3 they are 28\%, 38\%, and 34\%. The good agreement between the superposition of the calculated spectra and the experimental data reveals that the measured L$_{2,3}$-edge line shape can be accurately described as a quantitative combination of distinct oxidation-state contributions, thereby enabling their spatial differentiation, as shown in Figs.~\ref{fig:mn2plus}, \ref{fig:mn3plus}, \ref{fig:mn4plus}, and \ref{fig:lmoprofilemn}.

\begin{widetext}

\begin{figure}[!htbp]
    \centering
    \begin{subfigure}[b]{0.32\textwidth}
        \centering
        \includegraphics[width=\textwidth]{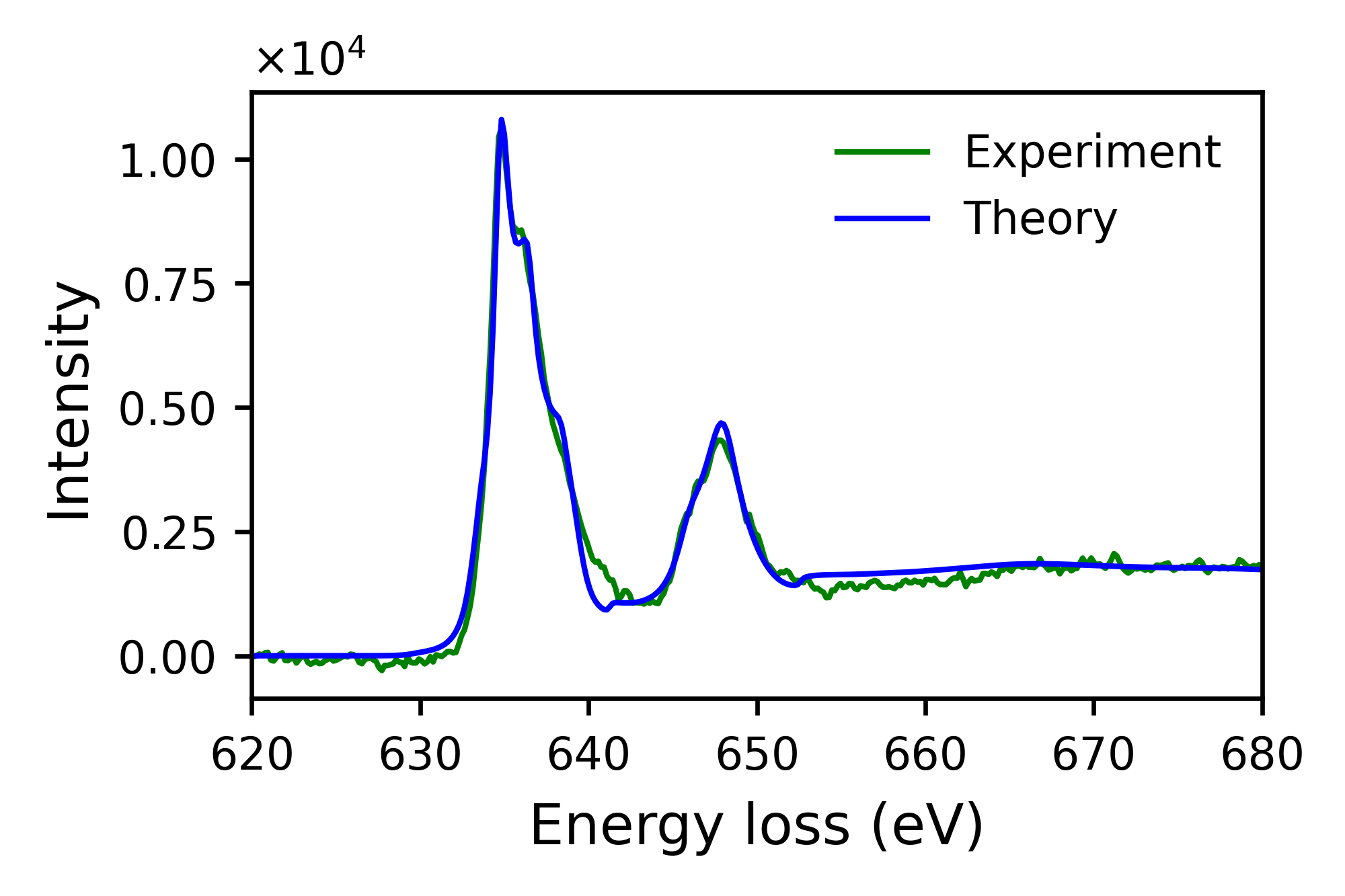}
        \caption{}
        \label{fig:Mnspecarea1}
    \end{subfigure}
    \hspace{0.1em}
    \begin{subfigure}[b]{0.32\textwidth}
        \centering
        \includegraphics[width=\textwidth]{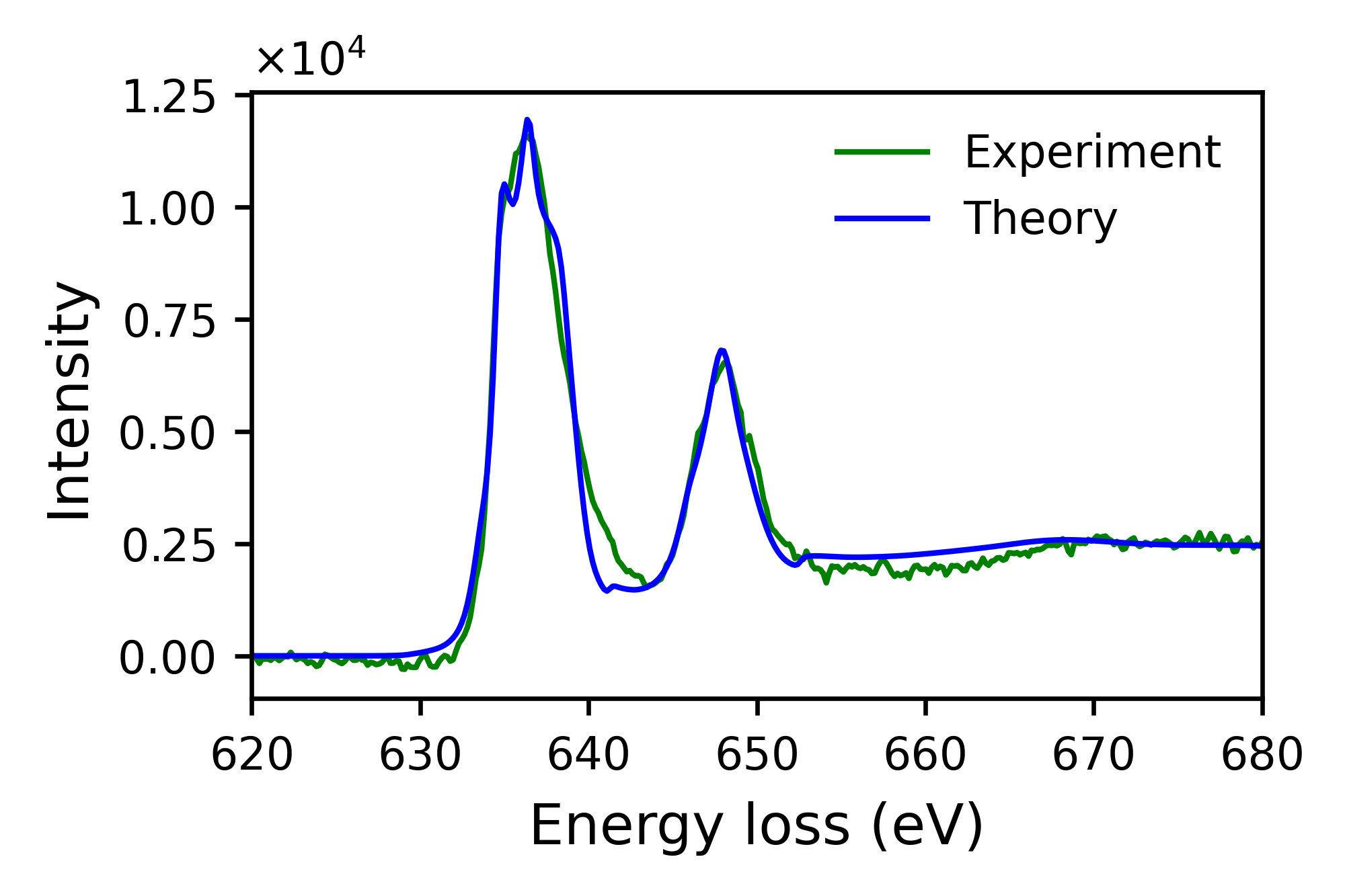}
        \caption{}
        \label{fig:Mnspecarea2}
    \end{subfigure}
    \hspace{0.1em}
    \begin{subfigure}[b]{0.32\textwidth}
        \centering
        \includegraphics[width=\textwidth]{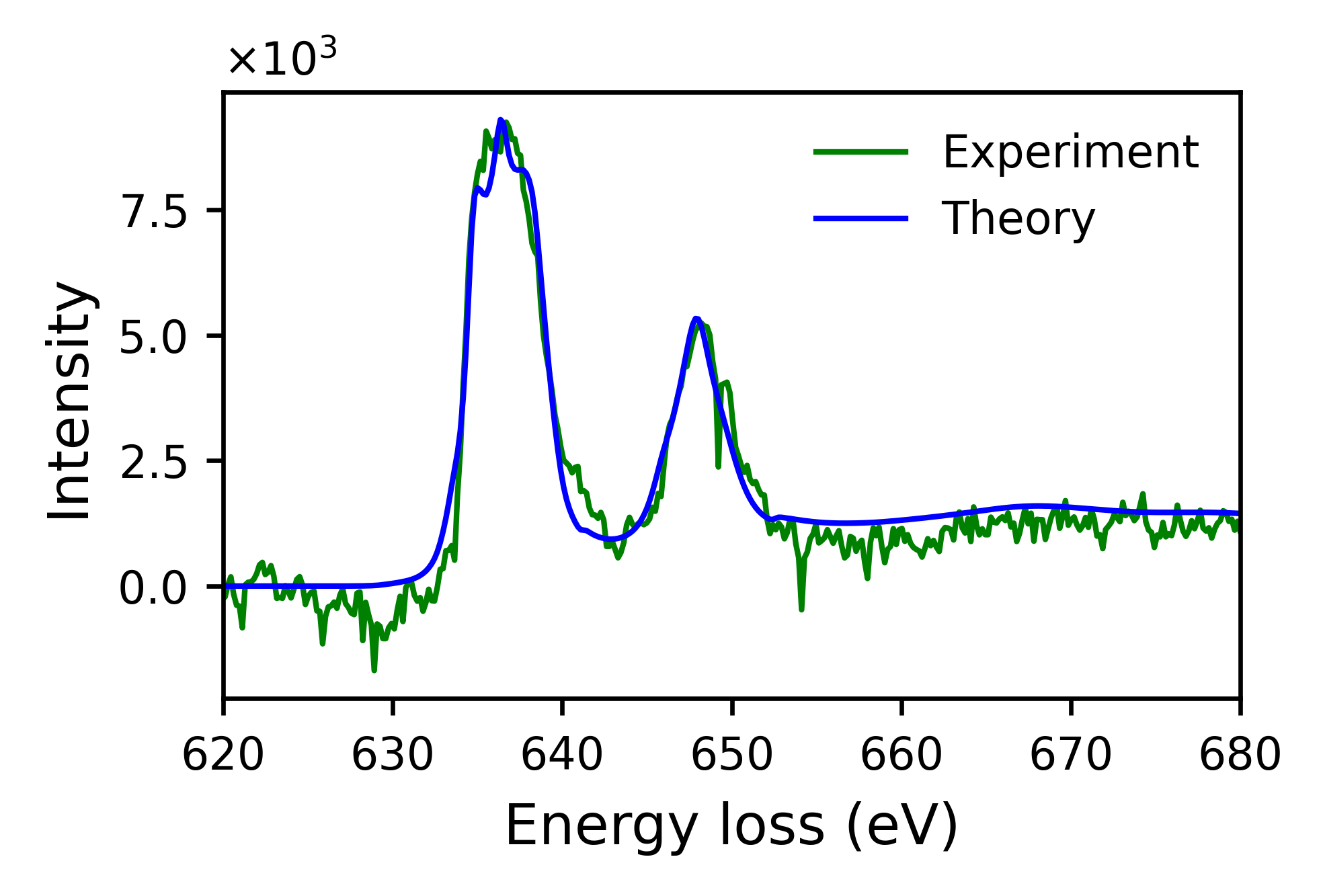}
        \caption{}
        \label{fig:Mnspecarea3}
    \end{subfigure}
    \begin{subfigure}[b]{0.32\textwidth}
        \centering
        \includegraphics[width=\textwidth]{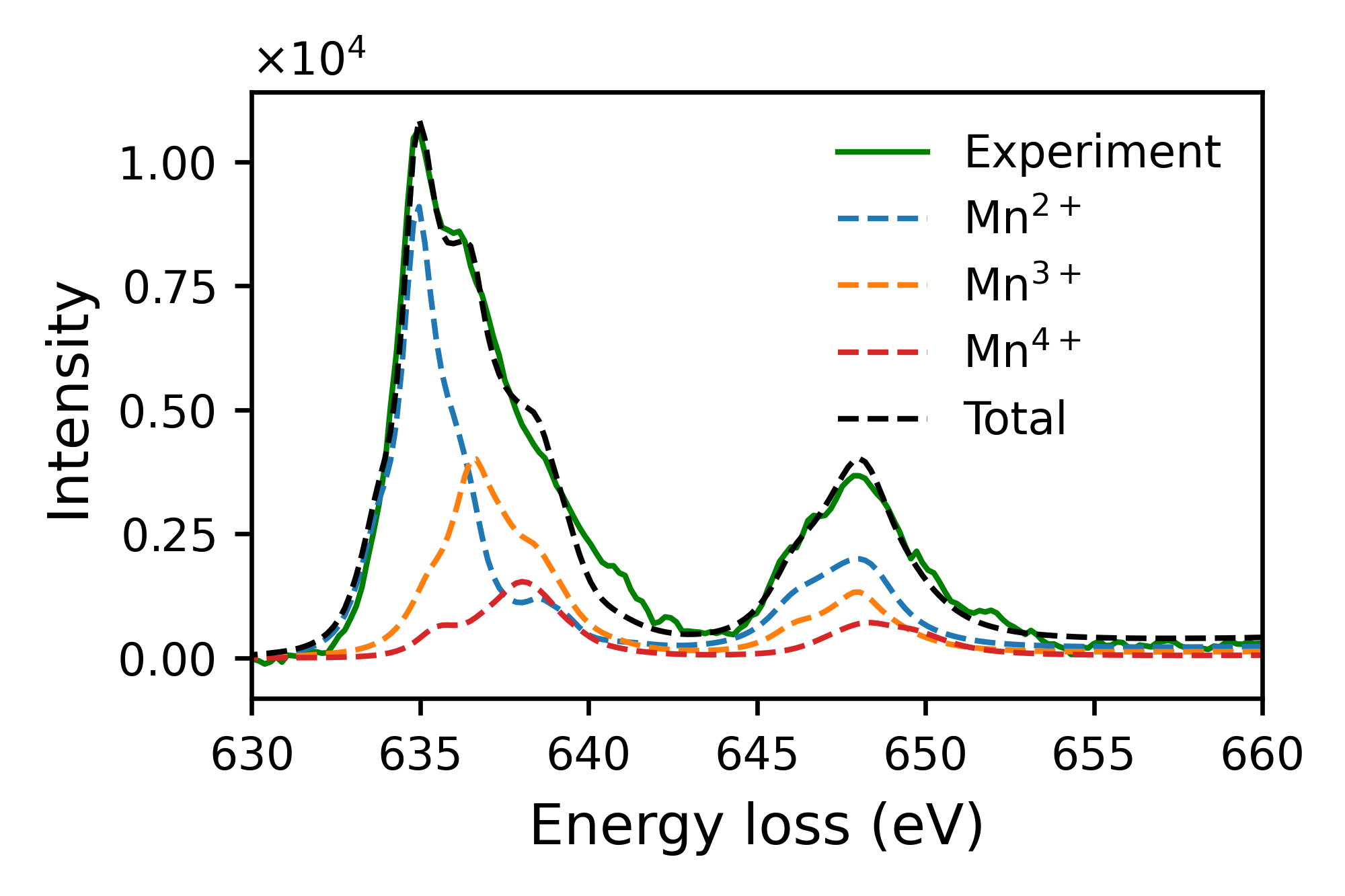}
        \caption{}
        \label{fig:Mnexcarea1}
    \end{subfigure}
    \hspace{0.1em}
    \begin{subfigure}[b]{0.32\textwidth}
        \centering
        \includegraphics[width=\textwidth]{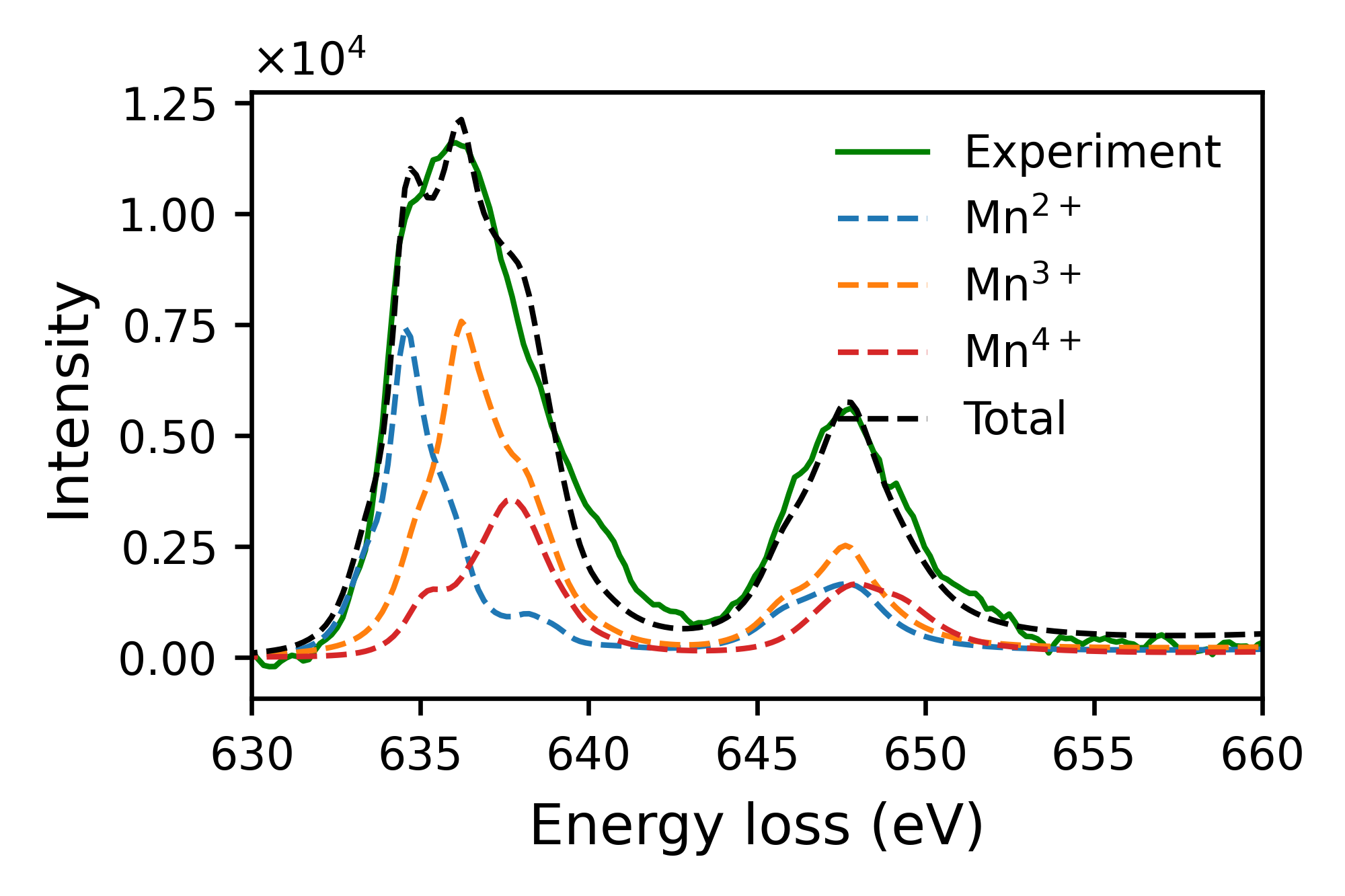}
        \caption{}
        \label{fig:Mnexcarea2}
    \end{subfigure}
    \hspace{0.1em}
    \begin{subfigure}[b]{0.32\textwidth}
        \centering
        \includegraphics[width=\textwidth]{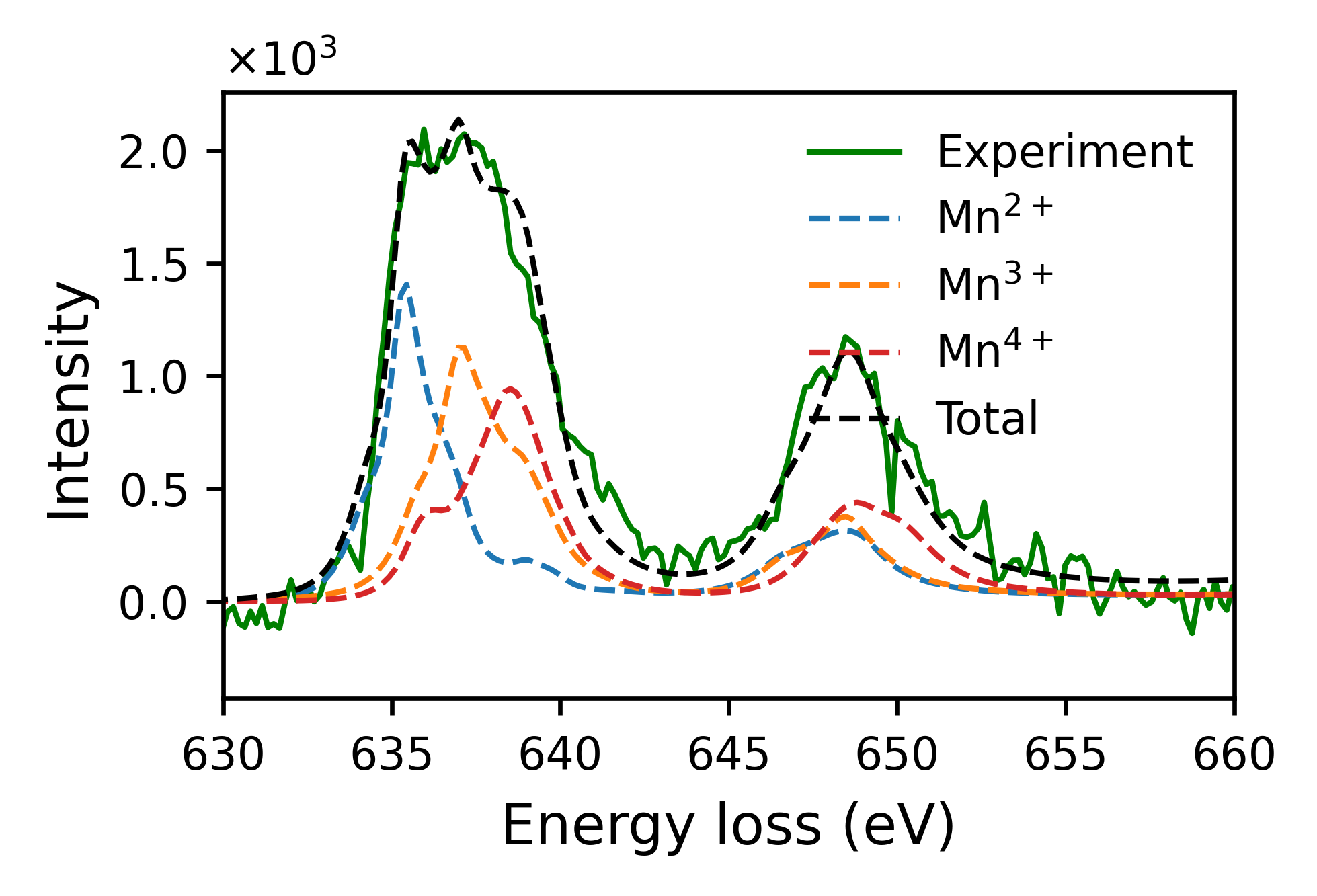}
        \caption{}
        \label{fig:Mnexcarea3}
    \end{subfigure}
    \caption{(a), (b), (c) compare the experimental spectra (green) with the calculated total inelastic differential cross section (blue) of Mn-L$_{2,3}$ from areas 1, 2, 3 as defined in Fig.~\ref{fig:dfslmo}, respectively. (d), (e), (f) depict the excitation part of spectra from areas 1, 2, 3, respectively. The green curve shows the experimental spectrum from which the background and the ionization differential cross section are subtracted. Only the discrete-excitation contribution is shown.}
    \label{fig:mnspeclmosto}
\end{figure}

\begin{figure}[!htbp]
%\centering
\begin{subfigure}[t]{0.33\linewidth}
\includegraphics[width=\linewidth]{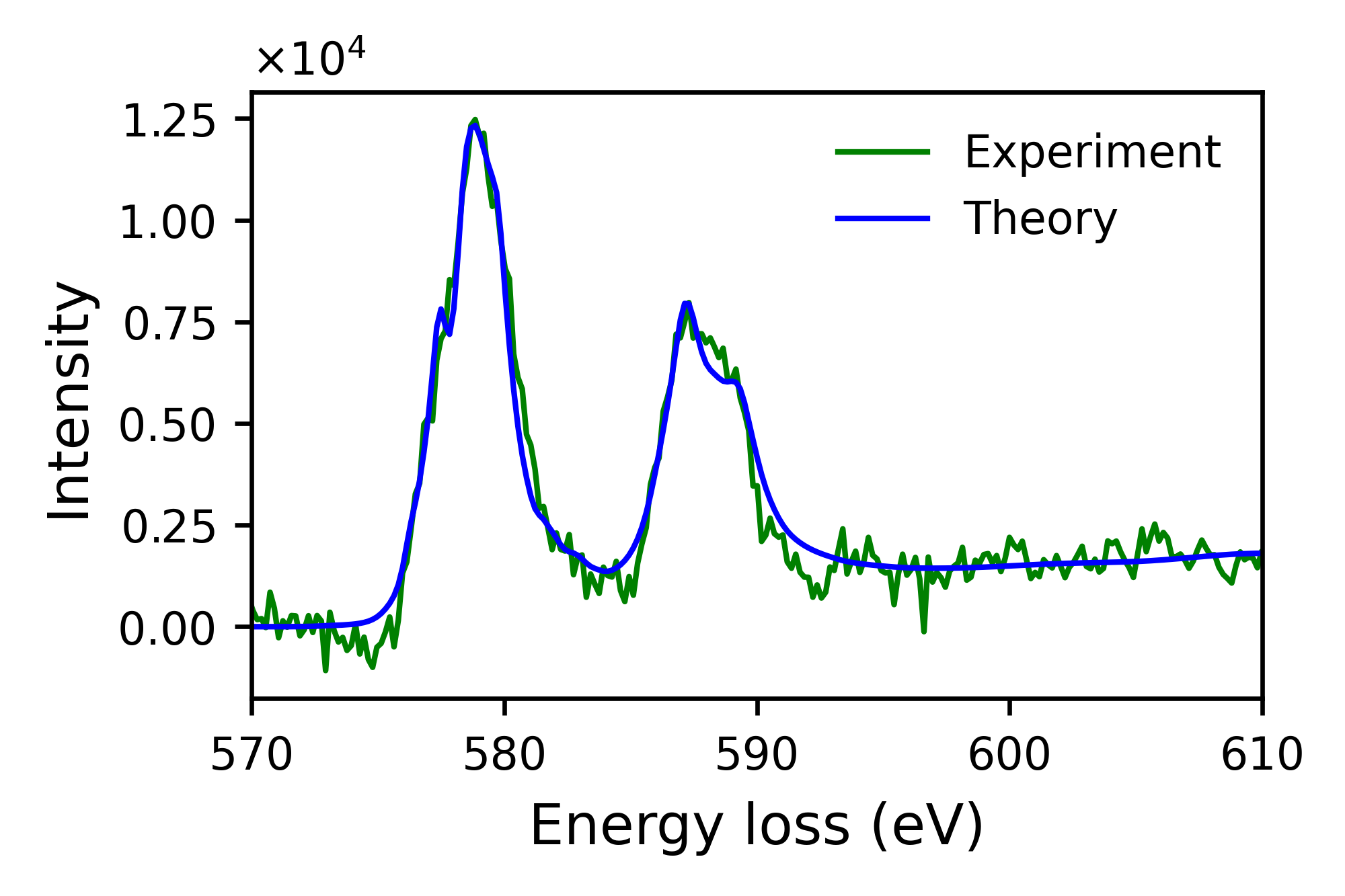}
\caption{}
\label{fig:Crspecarea3}
\end{subfigure}\hfill
\begin{subfigure}[t]{0.33\linewidth}
\includegraphics[width=\linewidth]{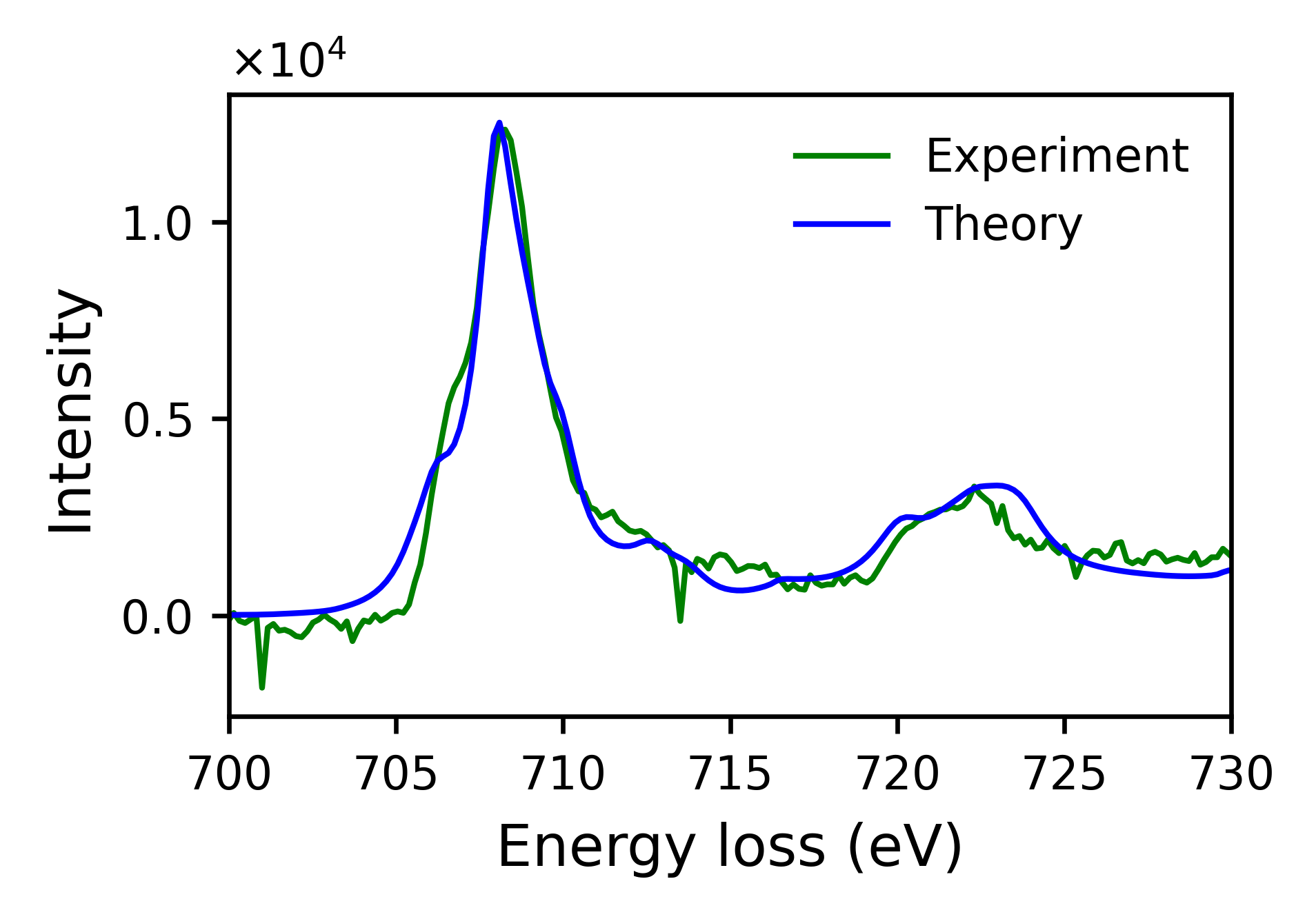}
\caption{}
\label{fig:Fespecarea3}
\end{subfigure}\hfill
\begin{subfigure}[t]{0.33\linewidth}
\includegraphics[width=\linewidth]{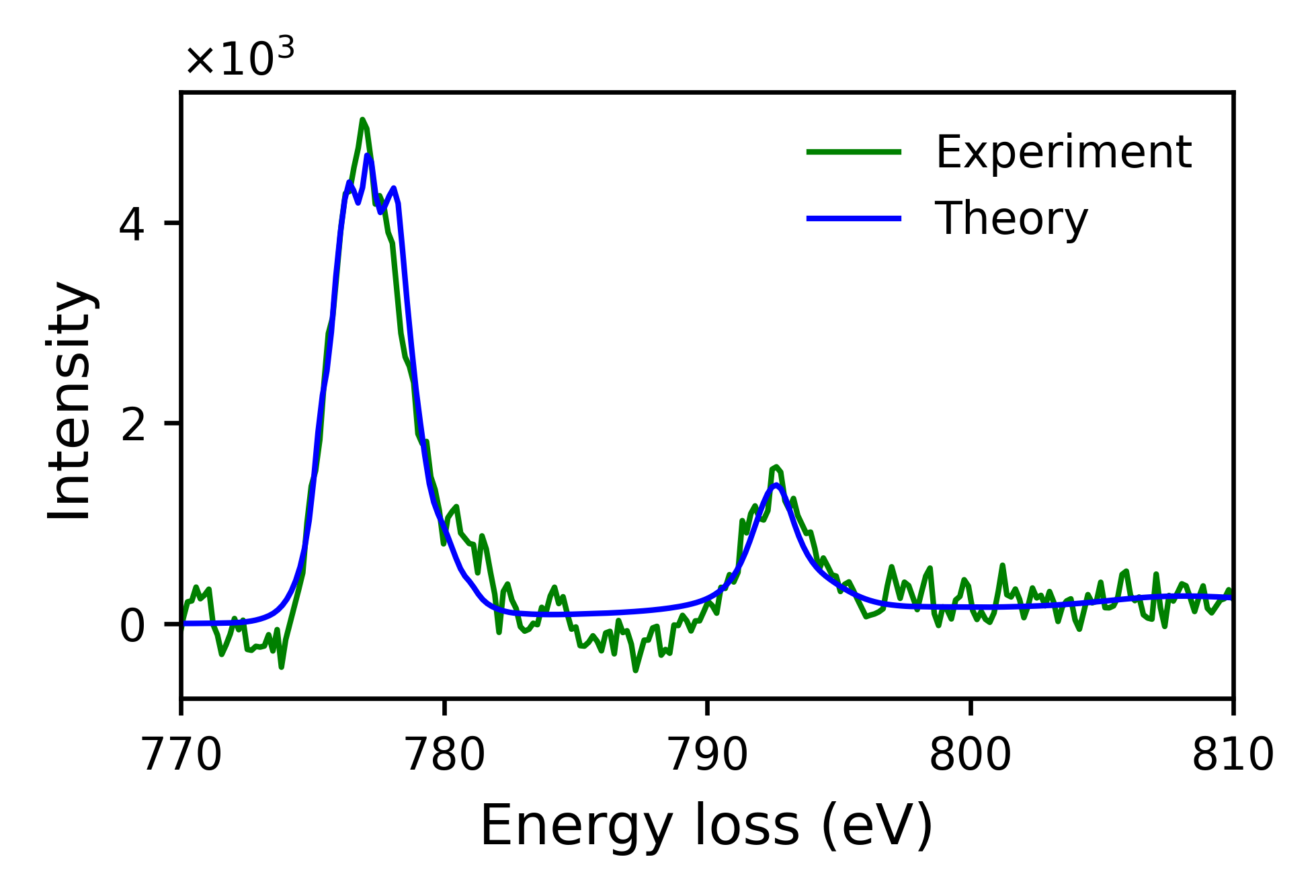}
\caption{}
\label{fig:Cospecarea3}
\end{subfigure}\hfill
\caption{(a), (b), (c) are comparisons between experiment (green) and calculated models (blue) at area 3 for $\mathrm{Cr}-L$, $\mathrm{Fe}-L$ and $\mathrm{Co}-L$ spectra, respectively.}
\label{fig:specarea3}
\end{figure}
\end{widetext}

\begin{figure}[!htbp] 
\begin{subfigure}[t]{.9\linewidth}
\caption{}
\includegraphics[width=\linewidth]{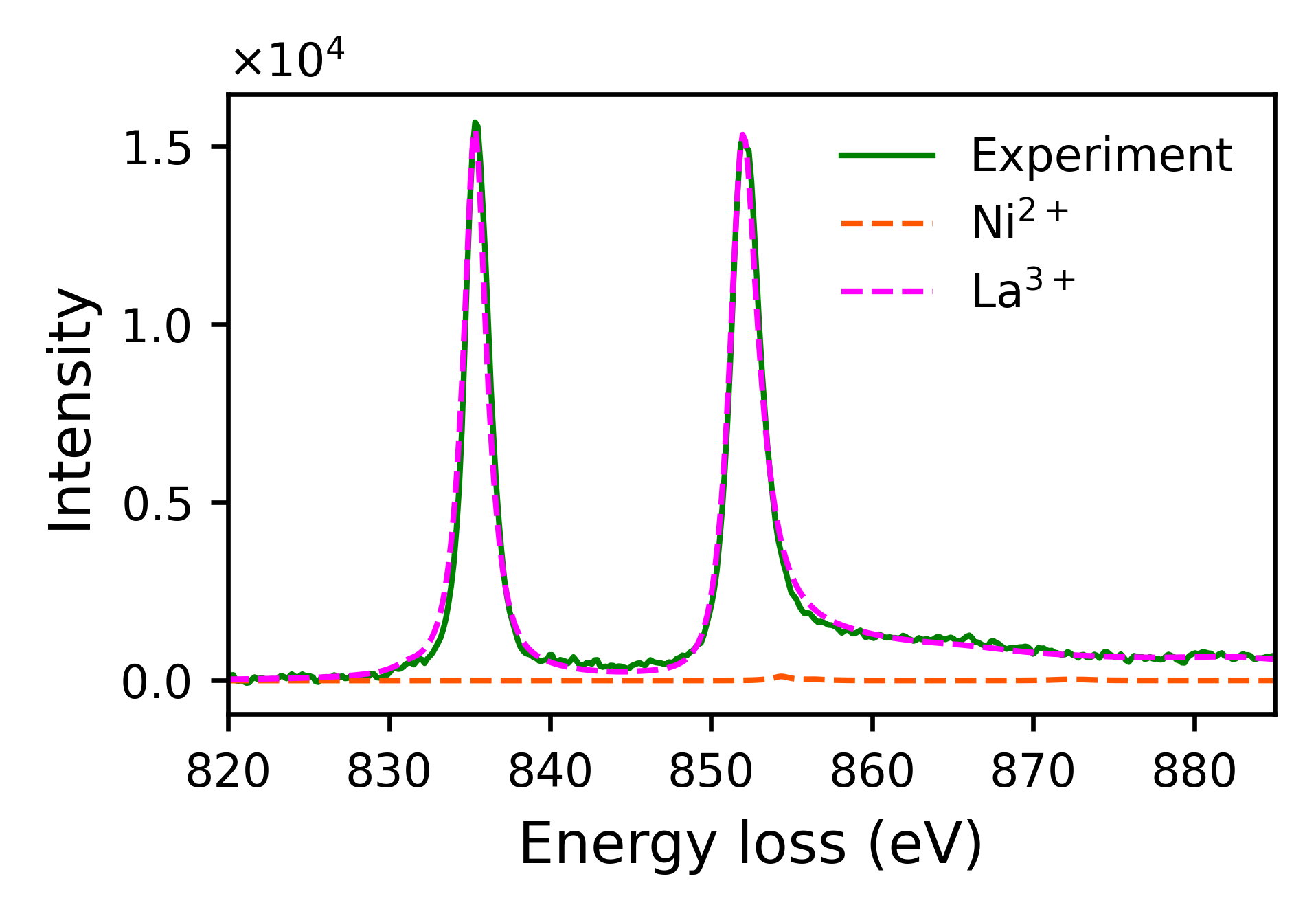}
\label{fig:area1LaNi}
\end{subfigure}\hfill%
\begin{subfigure}[t]{.9\linewidth}
\caption{}
\includegraphics[width=\linewidth]{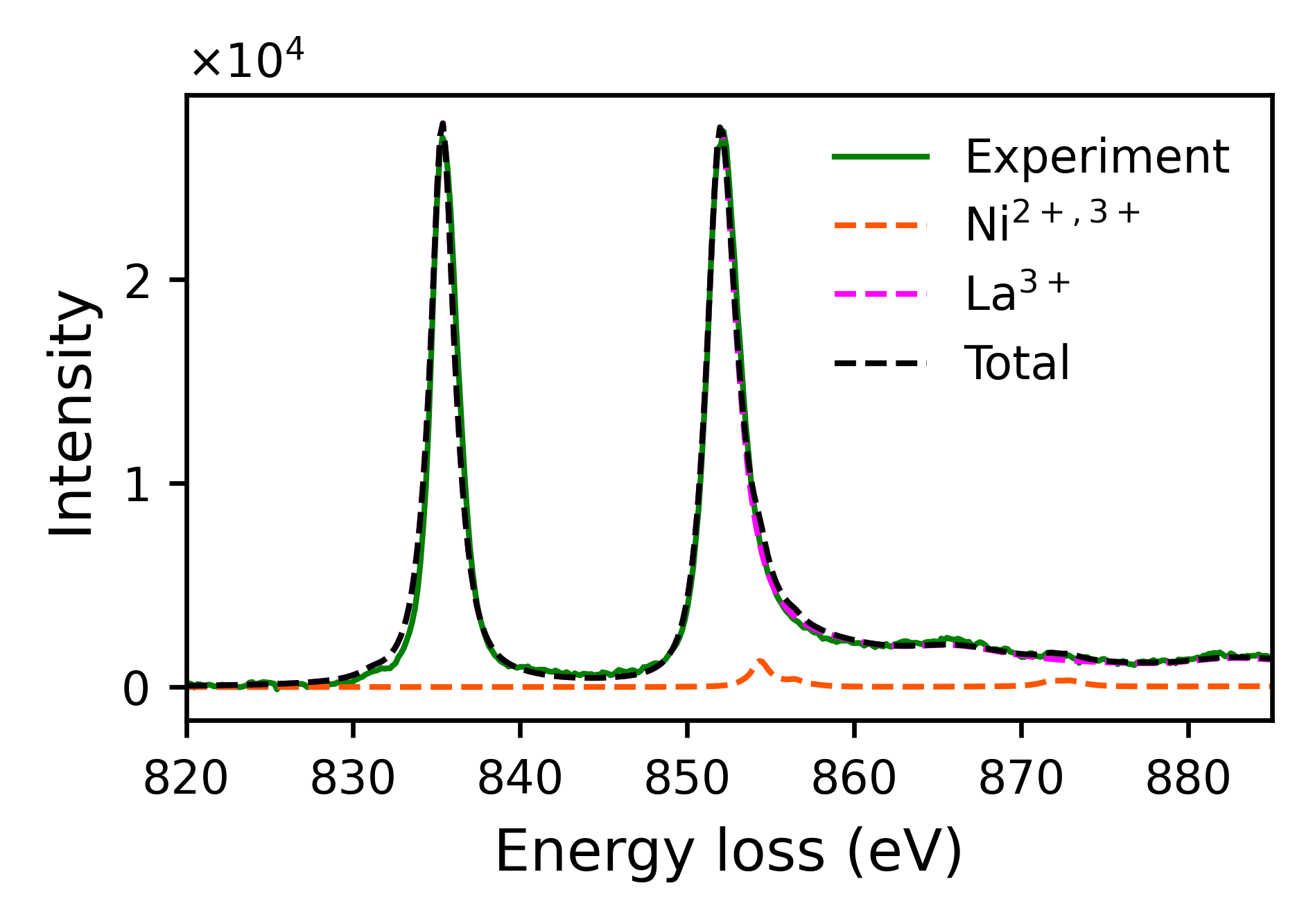}
\label{fig:area3LaNi}
\end{subfigure}%
\caption{The La$^{3+}$ (pink), the combined Ni$^{2+}$ and Ni$^{3+}$ (orange) spectral components and their sum (black) are compared with the experimental spectra (green) from areas 1 (a) and 3 (b), as defined in Fig.~\ref{fig:dfslmo}. The black curve is not shown in panel (a) since it essentially consists of the La$^{3+}$ model only. In area 1, the Ni contribution is negligible, whereas in area 3 a small Ni contribution is observed. The plotted Ni model is the sum of the Ni$^{2+}$ and Ni$^{3+}$ components. The background and ionization contributions have been subtracted from both the experimental data and the models.}
\label{fig:lsmoLaNispectra}
\end{figure}

When multiple transition metals occupy the $B$ site, as in LaCr$_{0.2}$Mn$_{0.2}$Fe$_{0.2}$Co$_{0.2}$Ni$_{0.2}$O$_3$ and in mixed metal  and high-entropy perovskites in general, charge neutrality does not generally imply a uniform $3+$ oxidation state across all cations, even in the absence of A-site doping. Instead, charge compensation occurs through selective stabilization of particular valence and spin states, driven by differences in crystal-field splitting, Hund's exchange, local bonding geometry, and metal-oxygen covalency \cite{Abbate1992,deGroot2005, doi:10.1021/acs.jpcc.6b00949}. In practice, Fe often remains in a comparatively robust high-spin trivalent state, whereas Mn accommodates mixed Mn$^{2+}$/Mn$^{3+}$/Mn$^{4+}$ populations, Ni and Co can adopt more flexible valence and spin states \cite{doi:10.1021/acs.jpcc.6b00949, PhysRevB.77.125124, kxnn-w1rk}, and Cr exhibits a predominantly 3+ oxidation state in the absence of doping- or environment-induced segregation effects \cite{wang2025selectiveoxidationcrsegregation}. Experimental studies using soft X-ray absorption and electron energy-loss spectroscopy have shown that cobalt in LaMn$_{1-x}$Co$_x$O$_3$-derived systems frequently stabilizes in a high-spin Co$^{2+}$ ($3d^7$) configuration over a broad range of compositions and thermodynamic conditions \cite{doi:10.1021/acs.jpcc.6b00949,Che2024}. In the present HEO layer, the Co population is predominantly Co$^{2+}$, while the mixed Mn population contains a substantial Mn$^{4+}$ contribution that partially compensates the Co$^{2+}$ charge deficit. The simultaneous Mn$^{2+}$ contribution, however, results in an average Mn valence close to 3+. As a result, the relevant electronic structure is governed by the local distribution of site-specific configurations rather than by a uniform oxidation state across the different $B$-site elements. Figs. \ref{fig:Crspecarea3}, \ref{fig:Fespecarea3} and \ref{fig:Cospecarea3} show good agreement between the calculated and experimental Cr$^{3+}$, Fe$^{3+}$ and Co$^{2+, 3+}$ spectra.  The Co$^{2+}$ and Co$^{3+}$ components are fitted to the Co edge in Fig. \ref{fig:Cospecarea3}, yielding relative populations of 95\% and 5\%, respectively, supporting Cr$^{3+}$, Fe$^{3+}$, together with predominantly Co$^{2+}$ character in the HEO layer. The Ni content in the HEO layer is also identified despite its much lower concentration than La and the strong overlap of their spectral features. The theoretical Ni$^{2+}$ and Ni$^{3+}$ spectral components are fitted to the experimental data, and their summed fitted contribution is presented in Figs.~\ref{fig:lmoprofileheo} and \ref{fig:lsmoLaNispectra}.  At sufficiently low concentrations, the precision of the Ni quantification remains limited by the Poisson counting statistics of the experimental data. For the profiles shown in Fig.~\ref{fig:lmoprofileheo}, the spectra were spatially filtered using a Gaussian kernel with $\sigma = 2.2$ pixels, instead of the $\sigma = 0.7$ used for the distribution maps, prior to fitting to improve the signal-to-noise ratio of the weak components.  The validity of the Ni spectral model and La/Ni deconvolution is further supported by the analysis of the NdNiO$_3$/La$_{0.67}$Sr$_{0.33}$MnO$_3$/NdGaO$_3$ sample in Fig. \ref{fig:lanispecnno} where the calculated Ni-L$_{2,3}$ spectral shape agrees with experimental spectra across the spectrum image.

\begin{figure}[!htbp] 
%\begin{subfigure}[t]{.8\linewidth}
%\caption{}
\includegraphics[width=\linewidth]{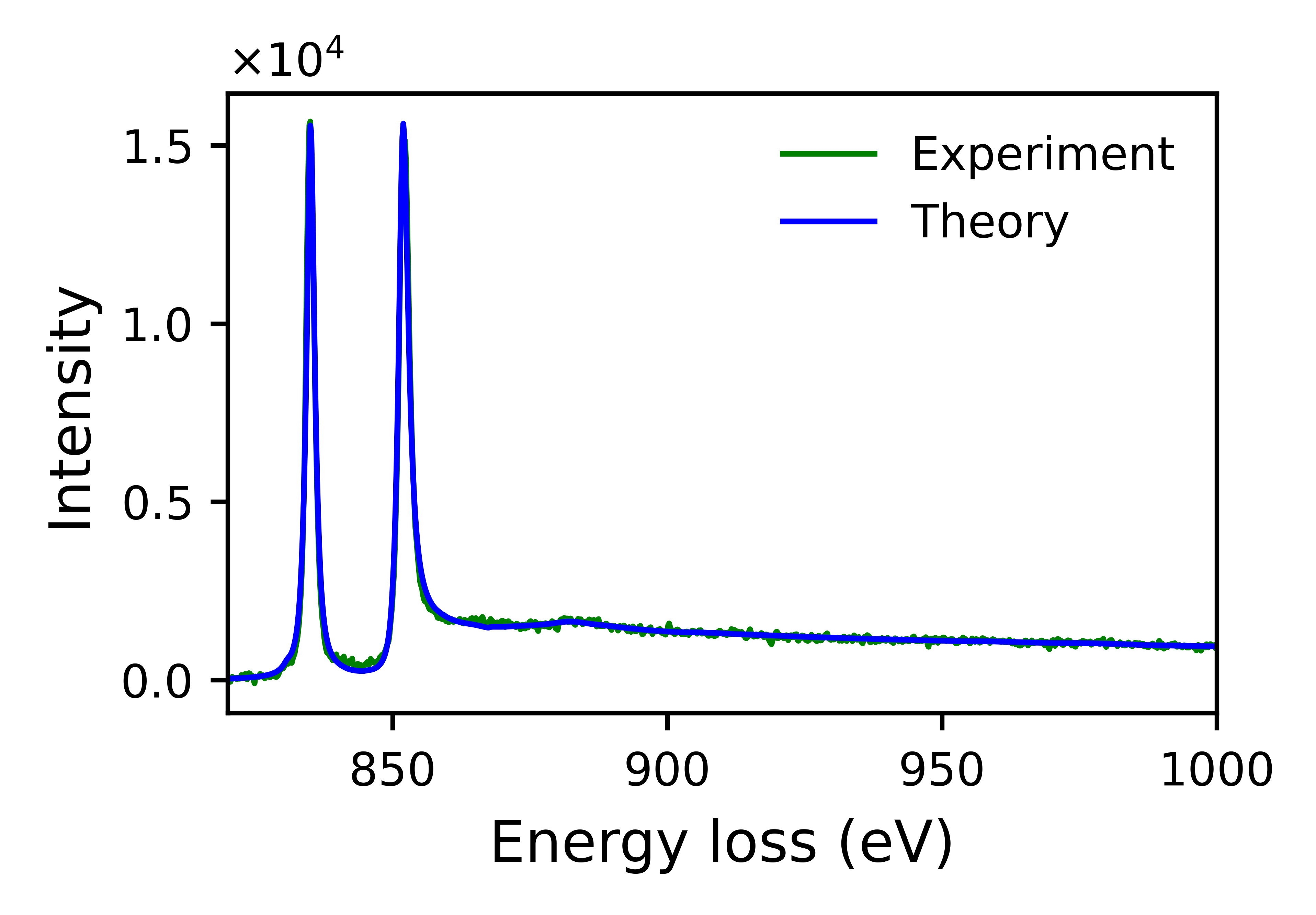}
\caption{The La-M$_{4,5}$ experimental spectrum (green) from area 1, as defined in Fig. \ref{fig:dfslmo}, compared to the theoretical inelastic cross section of La$^{3+}$(blue).}
\label{fig:Lalmosto}
\end{figure}

Good agreement of the theoretical inelastic cross section of La$^{3+}$ with the data is shown in Fig. \ref{fig:Lalmosto}, including the high-energy asymptotic behavior of the spectrum. Atomic correlation effects and the interaction of the discrete excitations with the continuum states  reproduce well the asymmetry of the experimental spectrum. The high-energy peaks which are observed in the spectrum are due to plural scattering which is appropriately accounted for by convolution with the low-loss spectrum. For rare-earth elements charge-transfer and hybridization effects are less pronounced  than for transition metals. The resulting agreement between theory and experiment reduces the model-related uncertainty in the quantitative determination of La and Ni, since the La M$_{4,5}$ and Ni L$_{2,3}$ edges strongly overlap, particularly at low Ni concentration.

\begin{figure}[!htbp] 
\includegraphics[width=\linewidth]{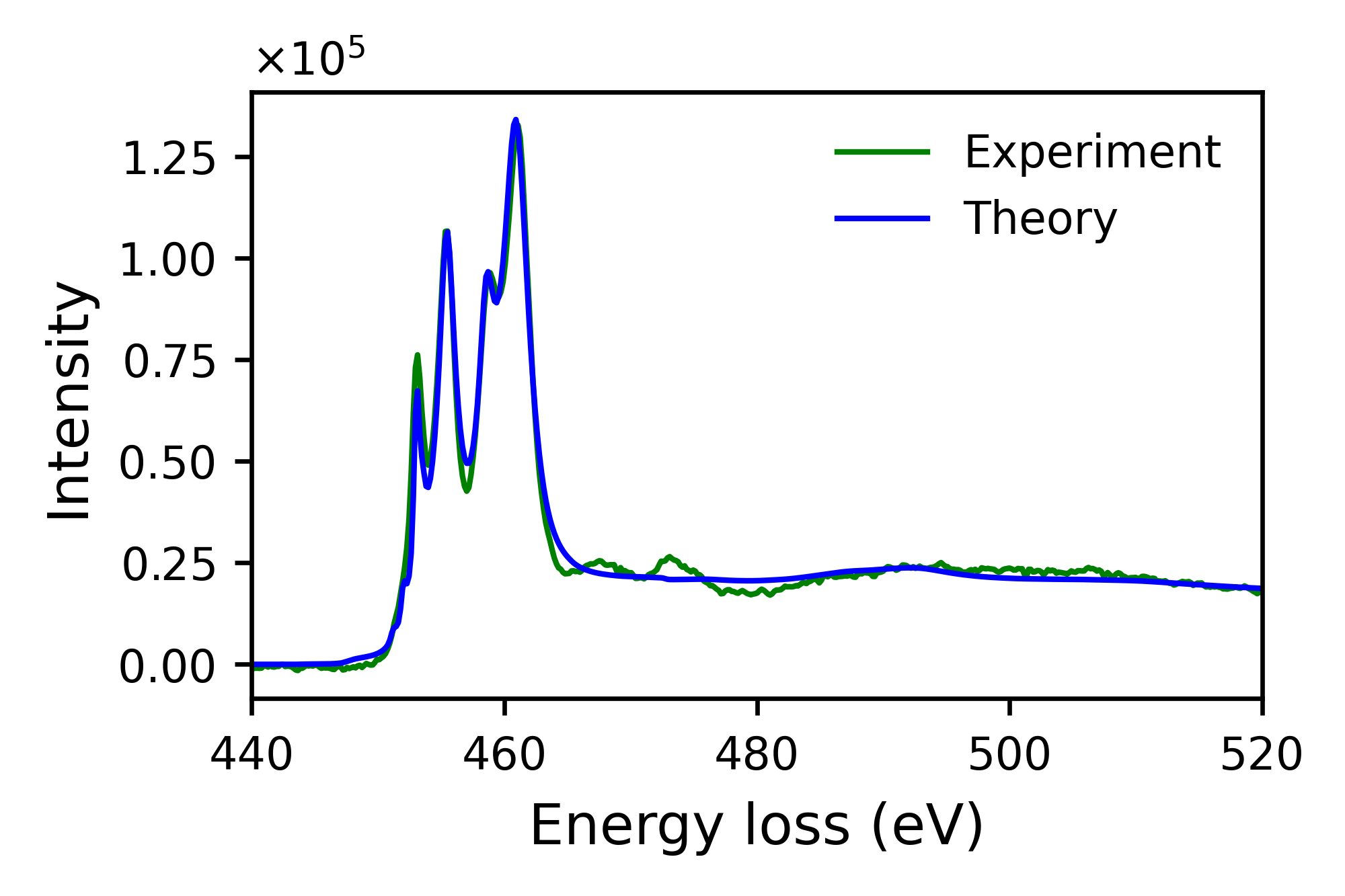}
\caption{The Ti-L spectrum from area 4 compared to the theoretical inelastic cross section for Ti$^{4+}$ in  octahedral symmetry.}
\label{fig:tispecarea4}
\end{figure}

The Ti$^{4+}$ signal is sharply confined to the SrTiO$_3$ substrate located at the bottom of the spectrum image in Fig. \ref{fig:ti4plus}, with no fitted evidence for significant Ti reduction in the present sample.  The Ti$^{4+}$ state is modeled in a 3d$^0$ configuration in $O_h$ symmetry \cite{deGroot1992}. Figure \ref{fig:tispecarea4} depicts  the fit of the theoretical spectral shape  to the spectrum from area 4 as defined in Fig. \ref{fig:dfslmo}. Overall, the EELS data support a picture of surface reduction, mixed-valence manganite behavior in both La$_{0.67}$Sr$_{0.33}$MnO$_3$ and the HEO film interior, relatively uniform Cr$^{3+}$, Fe$^{3+}$ and predominantly Co$^{2+}$ distributions throughout the HEO and a largely unreduced Ti$^{4+}$ substrate. The impact of plural scattering appears as a very broad peak at higher energy $\sim 490$ eV. Its detailed structure is modeled by the higher energy conduction band density of states of SrTiO$_3$ \cite{36kd-rnvd}. Beyond the plural scattering contribution, we observe the charge transfer satellite peaks at energies $ \sim 470$ eV and $\sim 475$ eV \cite{PhysRevB.100.075146, PhysRevB.91.121112}.

\subsection{NdNiO\texorpdfstring{$_3$}{3}/La\texorpdfstring{$_{0.67}$}{0.67}Sr\texorpdfstring{$_{0.33}$}{0.33}MnO\texorpdfstring{$_3$}{3}/NdGaO\texorpdfstring{$_3$}{3}}
\label{sec:nno}

\begin{figure}[!htbp] 
\includegraphics[width=.4\linewidth]{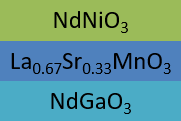}
\caption{A schematic view of the NdNiO$_3$/La$_{0.67}$Sr$_{0.33}$MnO$_3$/NdGaO$_3$ specimen.}
\label{fig:NNOschematic}
\end{figure}

NdNiO$_3$ provides another correlated-oxide reference system, Fig.~\ref{fig:NNOschematic}. Unlike LaNiO$_3$, bulk NdNiO$_3$ undergoes a metal-insulator transition accompanied by structural symmetry lowering and breathing distortions of the NiO$_6$ network, and its ground state is more appropriately viewed as strongly covalent or negative-charge-transfer in character than as a simple ionic Ni$^{3+}$ oxide \cite{PhysRevB.94.195127, Dom_nguez_2020}. In this picture, the insulating low-temperature phase involves inequivalent expanded and compressed NiO$_6$ octahedra with different degrees of Ni-O hybridization and ligand-hole character. Correspondingly, the Ni-L$_{2,3}$ fine structure is highly sensitive to the local electronic phase: near-atomic-scale STEM-EELS studies have shown that the O-K prepeak width and intensity track changes in Ni-O hybridization, while the Ni-L$_3$ edge develops a larger internal peak splitting and a shift of the lower-energy component in the insulating state \cite{doi:10.1021/acs.nanolett.0c04538}. In La-containing nickelate heterostructures, interpretation is further complicated by the strong overlap between the Ni-L$_{2,3}$ and La-M$_{4,5}$ edges, so quantitative recovery of the intrinsic Ni signal requires explicit deconvolution rather than empirical background subtraction \cite{PhysRevB.90.195140}.

\begin{figure}[!t] 
\includegraphics[width=\linewidth]{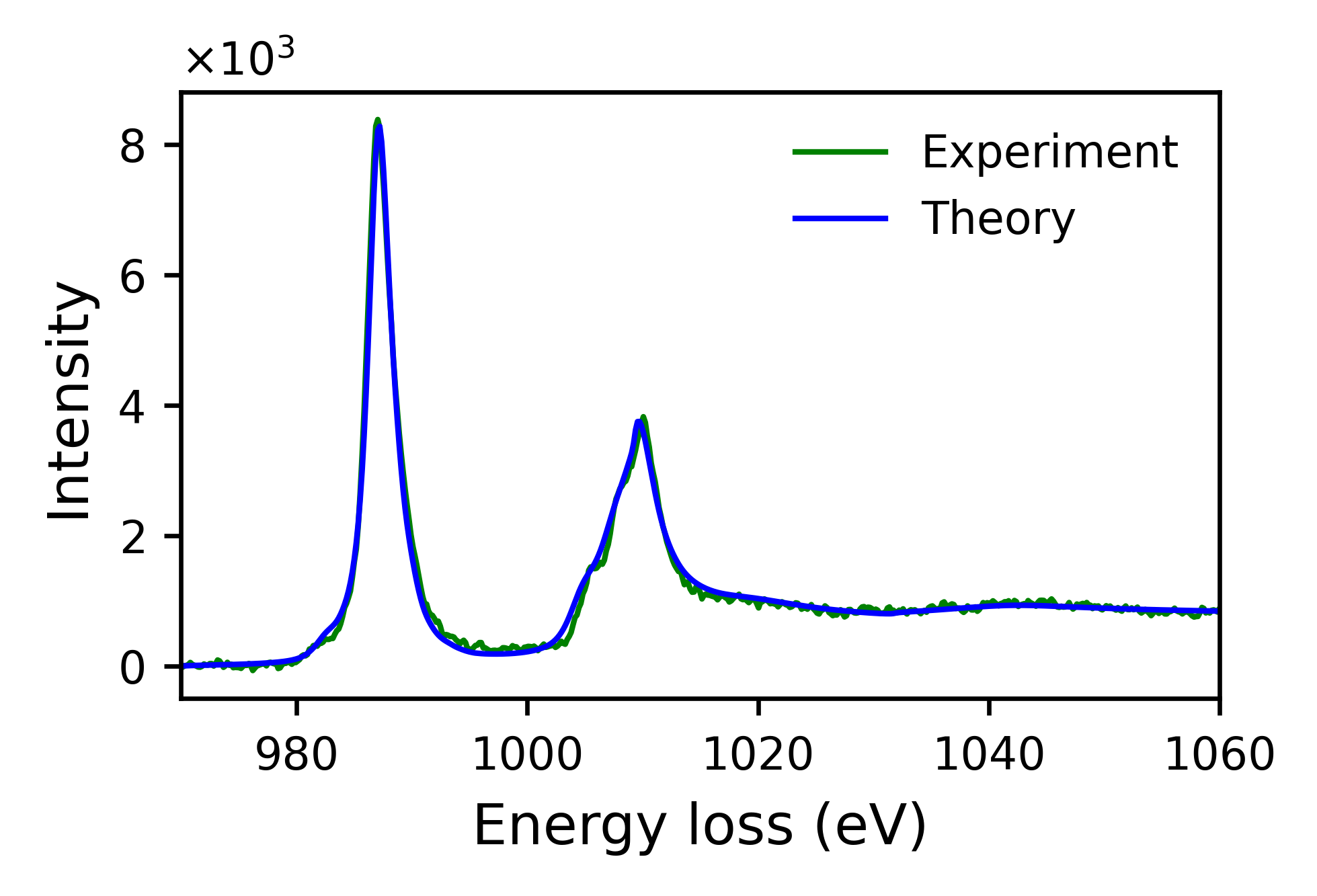}
\caption{The Nd-M$_{4,5}$ spectrum from area 1 (Fig.~\ref{fig:dfsnno}) compared to the theoretical inelastic cross section including the discrete and continuum excitations.} 
\label{fig:ndarea1}
\end{figure}

The Nd model, similarly to other rare-earth elements such as La, is governed by multiplet and correlation effects in the atomic structure. Fig.~\ref{fig:ndarea1} compares the calculated Nd$^{3+}$-M$_{4,5}$ spectral shape, including both discrete excitations and ionization contributions, with the experimental spectrum, showing good agreement between theory and experiment. The pronounced high-energy tails arise from the coupling of the discrete excitations to the continuum.

\begin{widetext}

\begin{figure}[!htbp]
    \centering
    \begin{subfigure}[b]{0.11\textwidth}
        \centering
        \includegraphics[width=.5\textwidth, keepaspectratio]{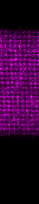}
        \caption{}
        \label{fig:Lanno}
    \end{subfigure}
    \hspace{0.1em}
    \begin{subfigure}[b]{0.11\textwidth}
        \centering
        \includegraphics[width=.5\textwidth, keepaspectratio]{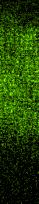}
        \caption{}
        \label{fig:mn2plusnno}
    \end{subfigure}
    \hspace{0.1em}
    \begin{subfigure}[b]{0.11\textwidth}
        \centering
        \includegraphics[width=.5\textwidth, keepaspectratio]{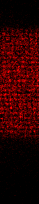}
        \caption{}
        \label{fig:mn3plusnno}
    \end{subfigure}
    \begin{subfigure}[b]{0.11\textwidth}
        \centering
        \includegraphics[width=.5\textwidth, keepaspectratio]{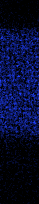}
        \caption{}
        \label{fig:mn4plusnno}
    \end{subfigure}
    \begin{subfigure}[b]{0.11\textwidth}
        \centering
        \includegraphics[width=.5\textwidth, keepaspectratio]{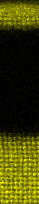}
        \caption{}
        \label{fig:nd}
    \end{subfigure}
    \begin{subfigure}[b]{0.11\textwidth}
        \centering
        \includegraphics[width=.5\textwidth, keepaspectratio]{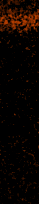}
        \caption{}
        \label{fig:ni2nno}
    \end{subfigure}
    \begin{subfigure}[b]{0.11\textwidth}
        \centering
        \includegraphics[width=.5\textwidth, keepaspectratio]{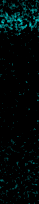}
        \caption{}
        \label{fig:ni3nno}
    \end{subfigure}
    \begin{subfigure}[b]{0.11\textwidth}
        \centering
        \includegraphics[width=.5\textwidth, keepaspectratio]{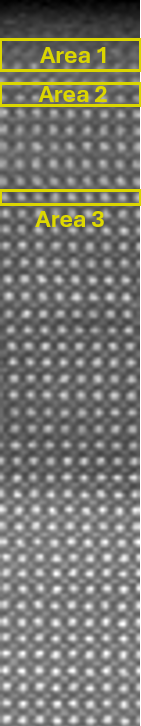}
        \caption{}
        \label{fig:dfsnno}
    \end{subfigure}
    \caption{The distribution maps of the intensity of (a) La$^{3+}$, (b) Mn$^{2+}$, (c) Mn$^{3+}$, (d) Mn$^{4+}$, (e) Nd$^{3+}$, (f) Ni$^{2+}$, (g) Ni$^{3+}$, (h) dark-field STEM image showing the integration areas from which the spectra used for further analysis are extracted. The top of the spectrum image corresponds to NdNiO$_3$, the middle layer is the La$_{0.67}$Sr$_{0.33}$MnO$_3$ while at the bottom there is the NdGaO$_3$ substrate.}
    \label{fig:MapsNd}
\end{figure}

\end{widetext}

In Fig.~\ref{fig:MapsNd}, near-atomic-resolution STEM-EELS maps of the NdNiO$_3$/La$_{0.67}$Sr$_{0.33}$MnO$_3$/NdGaO$_3$ heterostructure reveal a chemically well-defined interface together with clear spatial separation of the Nd/Ni-rich and La/Mn-rich regions. Within the La$_{0.67}$Sr$_{0.33}$MnO$_3$ layer, the Mn$^{2+}$, Mn$^{3+}$, and Mn$^{4+}$ component maps demonstrate mixed valence at near-atomic resolution, while on the nickelate side the recovered Ni$^{2+}$ and Ni$^{3+}$ components remain localized in the Nd-rich region after deconvolution from the overlapping La-M$_{4,5}$ edge. We observe limited La and Nd interdiffusion across the interfaces, although the individual layers remain clearly distinguishable. The maps also indicate a Mn$^{2+}$ contribution extending into the NdNiO$_3$ side, consistent with limited interfacial cation interdiffusion, but the dominant contrast remains that of a chemically well-defined manganite/nickelate interface.

\begin{widetext}

\begin{figure}[!h]
    \centering
    \begin{subfigure}[b]{0.32\textwidth}
        \centering
        \includegraphics[width=\textwidth]{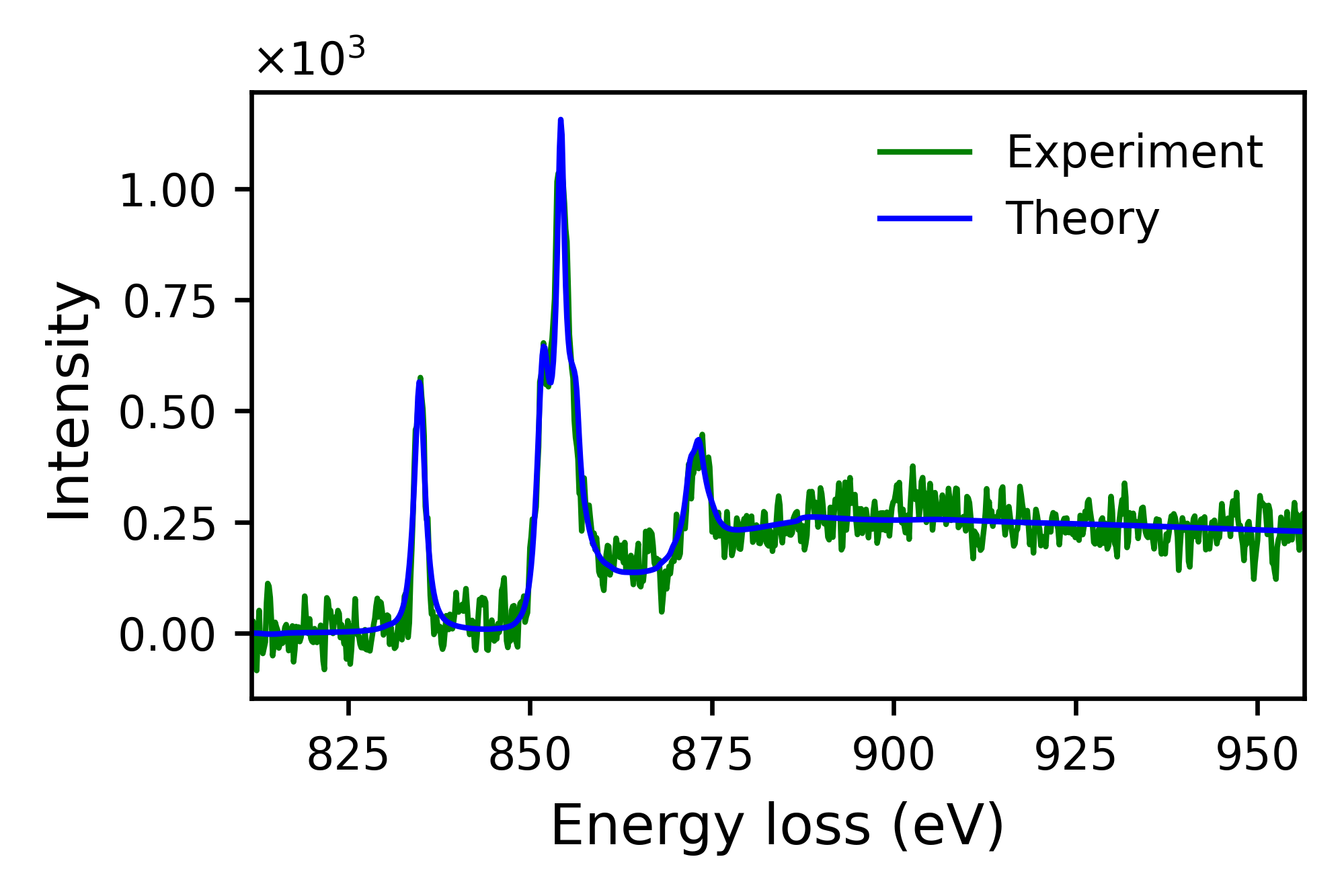}
        \caption{}
        \label{fig:nnoLaNispecarea1}
    \end{subfigure}
    \hspace{0.1em}
    \begin{subfigure}[b]{0.32\textwidth}
        \centering
        \includegraphics[width=\textwidth]{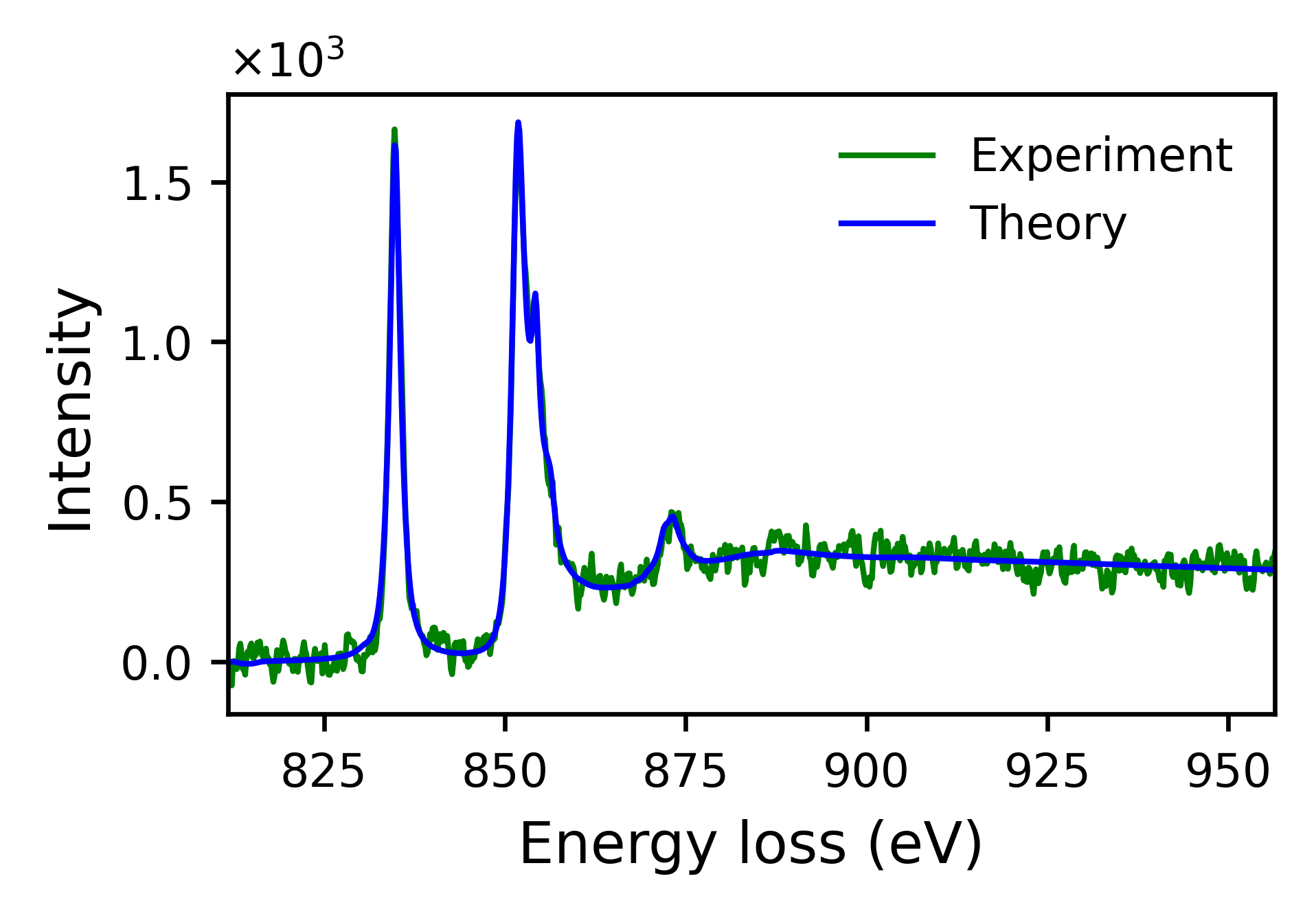}
        \caption{}
        \label{fig:nnoLaNispecarea2}
    \end{subfigure}
    \hspace{0.1em}
    \begin{subfigure}[b]{0.32\textwidth}
        \centering
        \includegraphics[width=\textwidth]{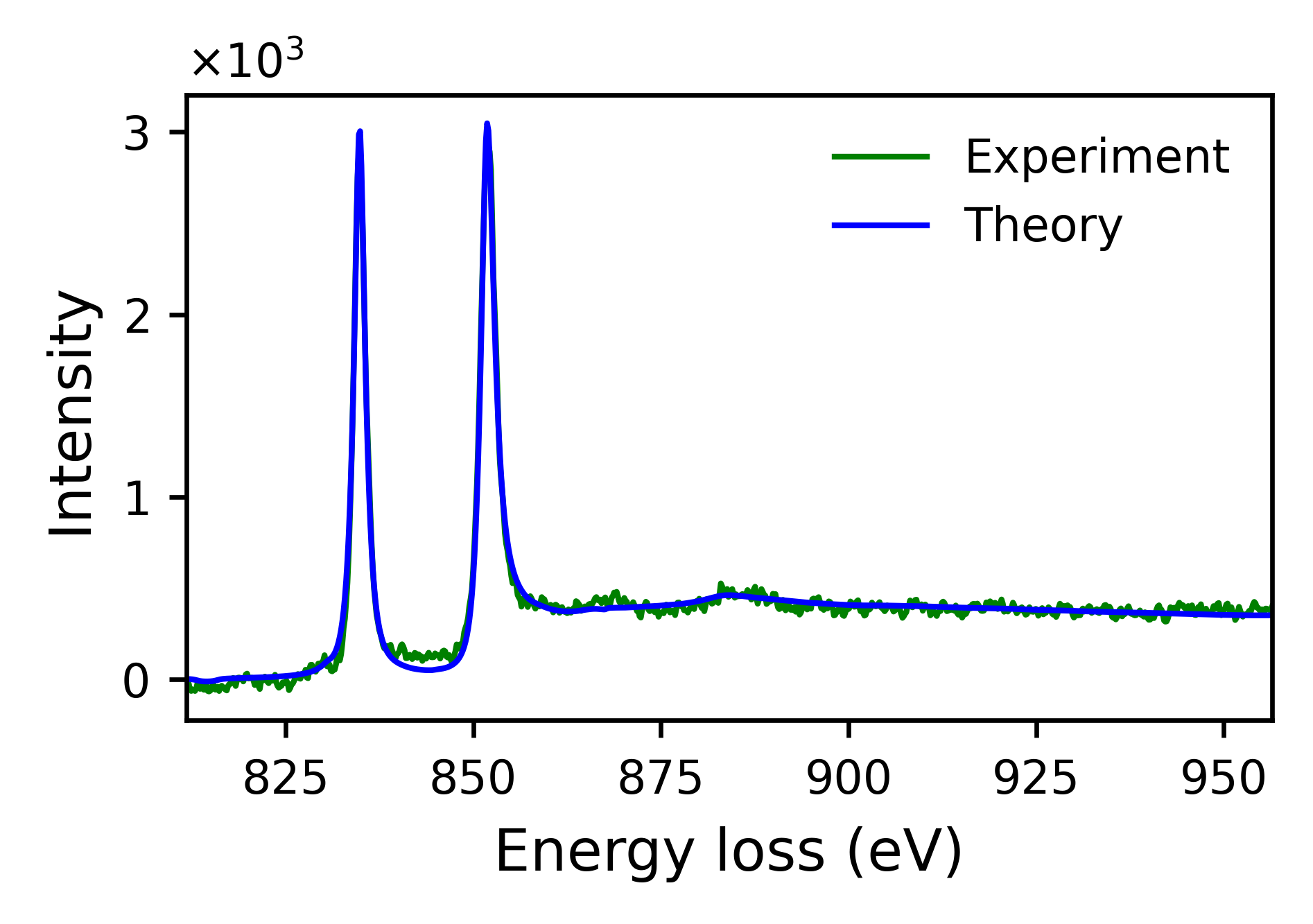}
        \caption{}
        \label{fig:nnoLaNispecarea3}
    \end{subfigure}
    \begin{subfigure}[b]{0.32\textwidth}
        \centering
        \includegraphics[width=\textwidth]{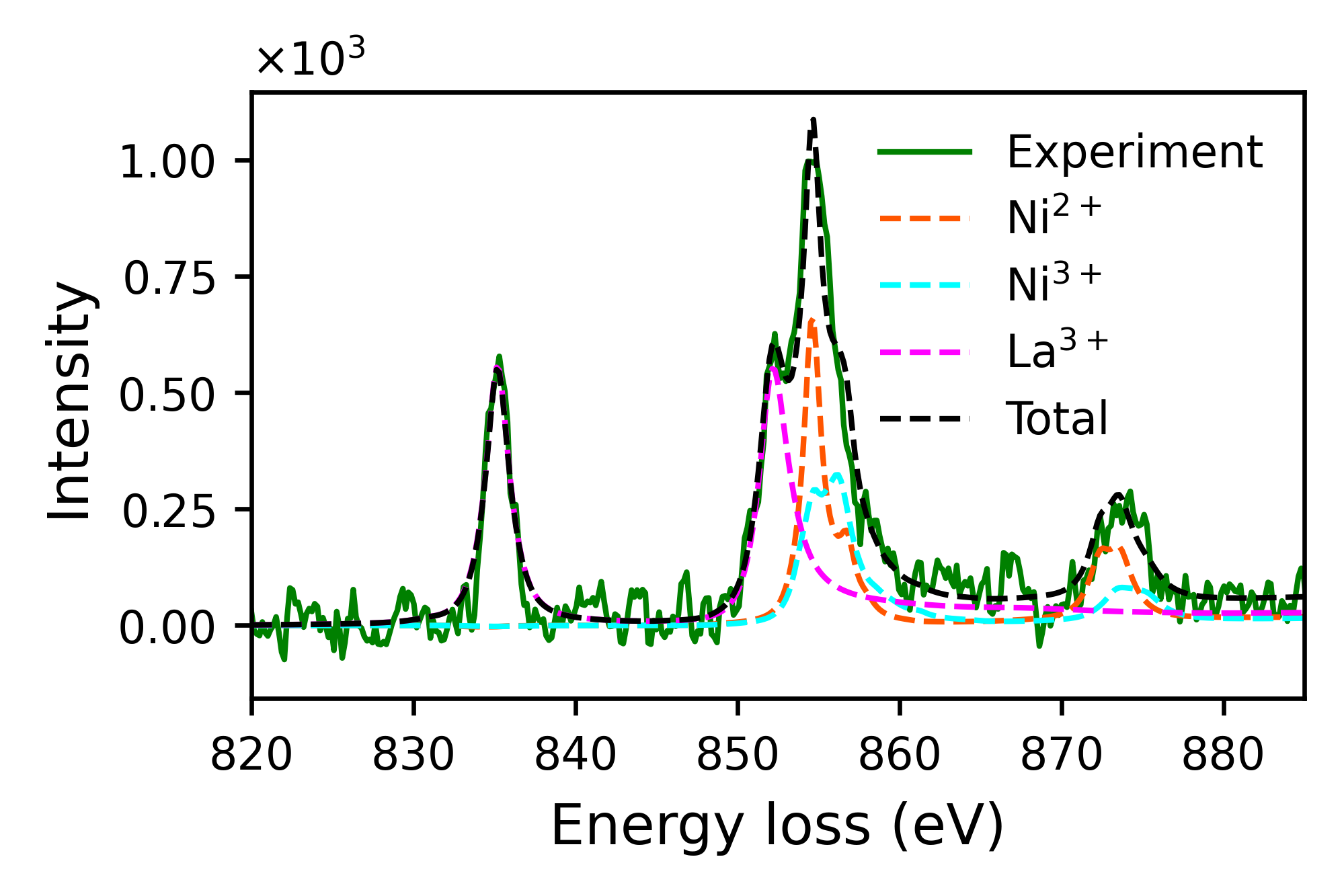}
        \caption{}
        \label{fig:laniexcarea1}
    \end{subfigure}
    \hspace{0.1em}
    \begin{subfigure}[b]{0.32\textwidth}
        \centering
        \includegraphics[width=\textwidth]{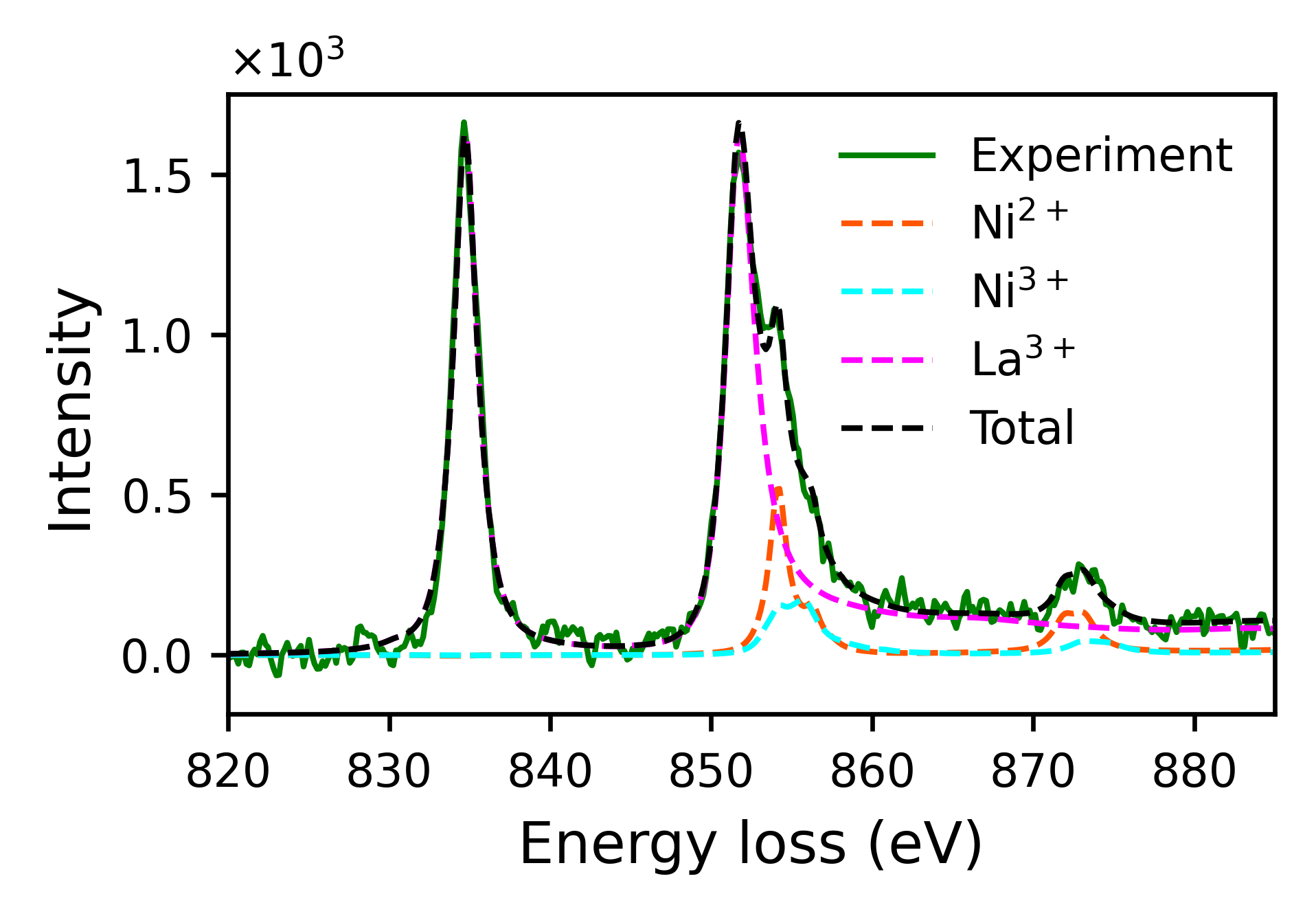}
        \caption{}
        \label{fig:laniexcarea2}
    \end{subfigure}
    \hspace{0.1em}
    \begin{subfigure}[b]{0.32\textwidth}
        \centering
        \includegraphics[width=\textwidth]{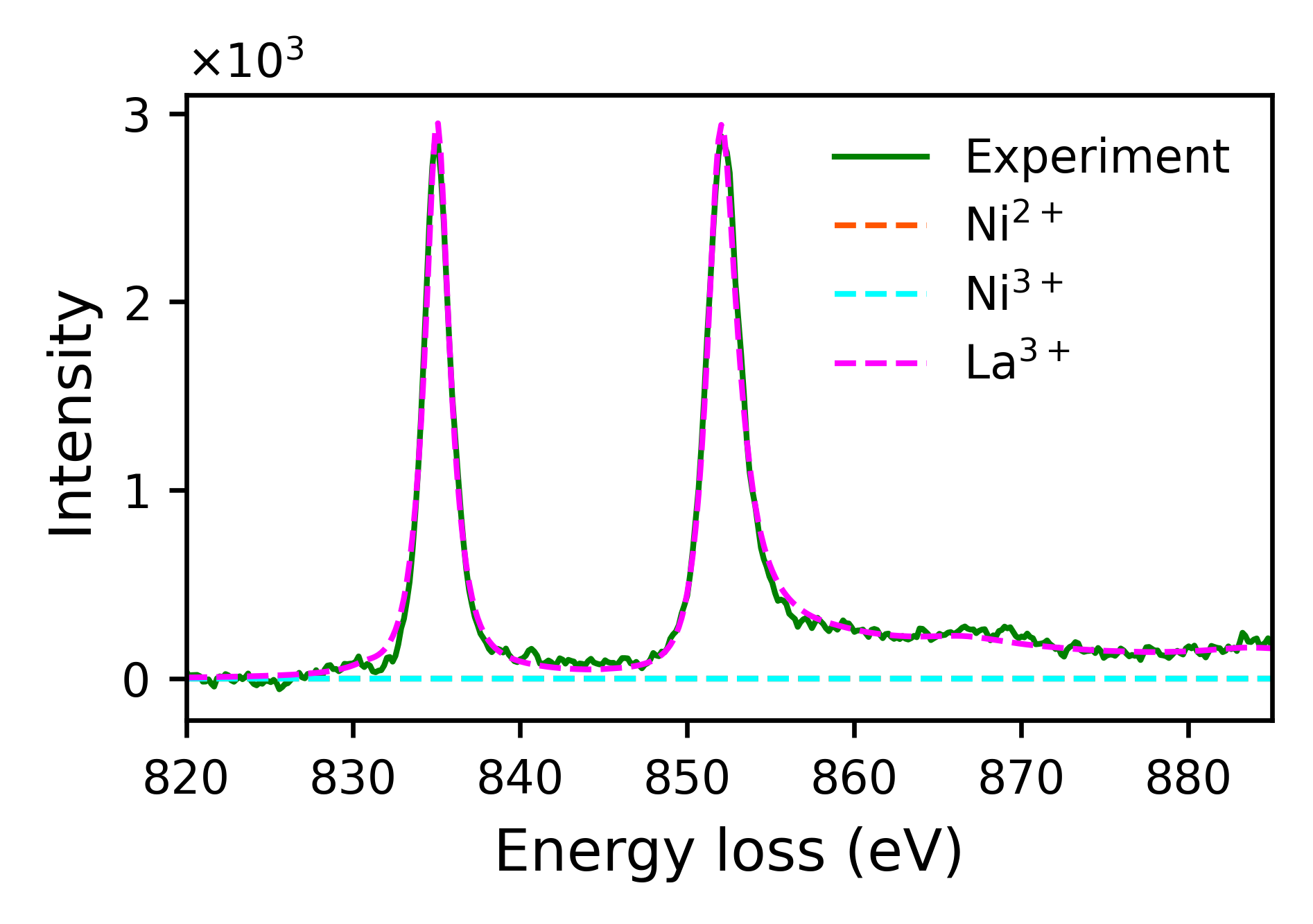}
        \caption{}
        \label{fig:laniexcarea3}
    \end{subfigure}
    \caption{(a), (b), (c) show comparisons between experiment (green) and calculated inelastic cross sections (blue) of Ni-L$_{2,3}$ and La-M$_{4,5}$ overlapping spectra at areas 1, 2, 3, as defined in Fig.~\ref{fig:dfsnno}. The calculated La and Ni spectra closely follow the data and enable the deconvolution of their respective spectral contributions and electronic states. In (d), (e) and (f), the individual components of the excitation parts of the spectra are compared to the experimental spectra from areas 1, 2 and 3 respectively after the background and ionization cross section subtraction. In (f) we did not plot the total cross section because the fitted Ni contribution is essentially zero.}
    \label{fig:lanispecnno}
\end{figure}

\end{widetext}

\begin{figure}[h] 
\begin{subfigure}[t]{.8\linewidth}
\caption{}
\includegraphics[width=\linewidth]{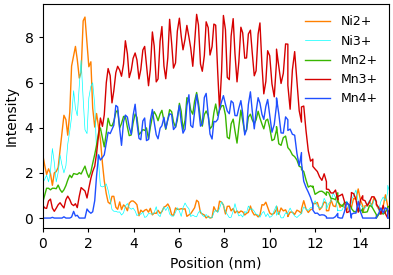}
\end{subfigure}\hfill%
\begin{subfigure}[t]{.8\linewidth}
\caption{}
\includegraphics[width=\linewidth]{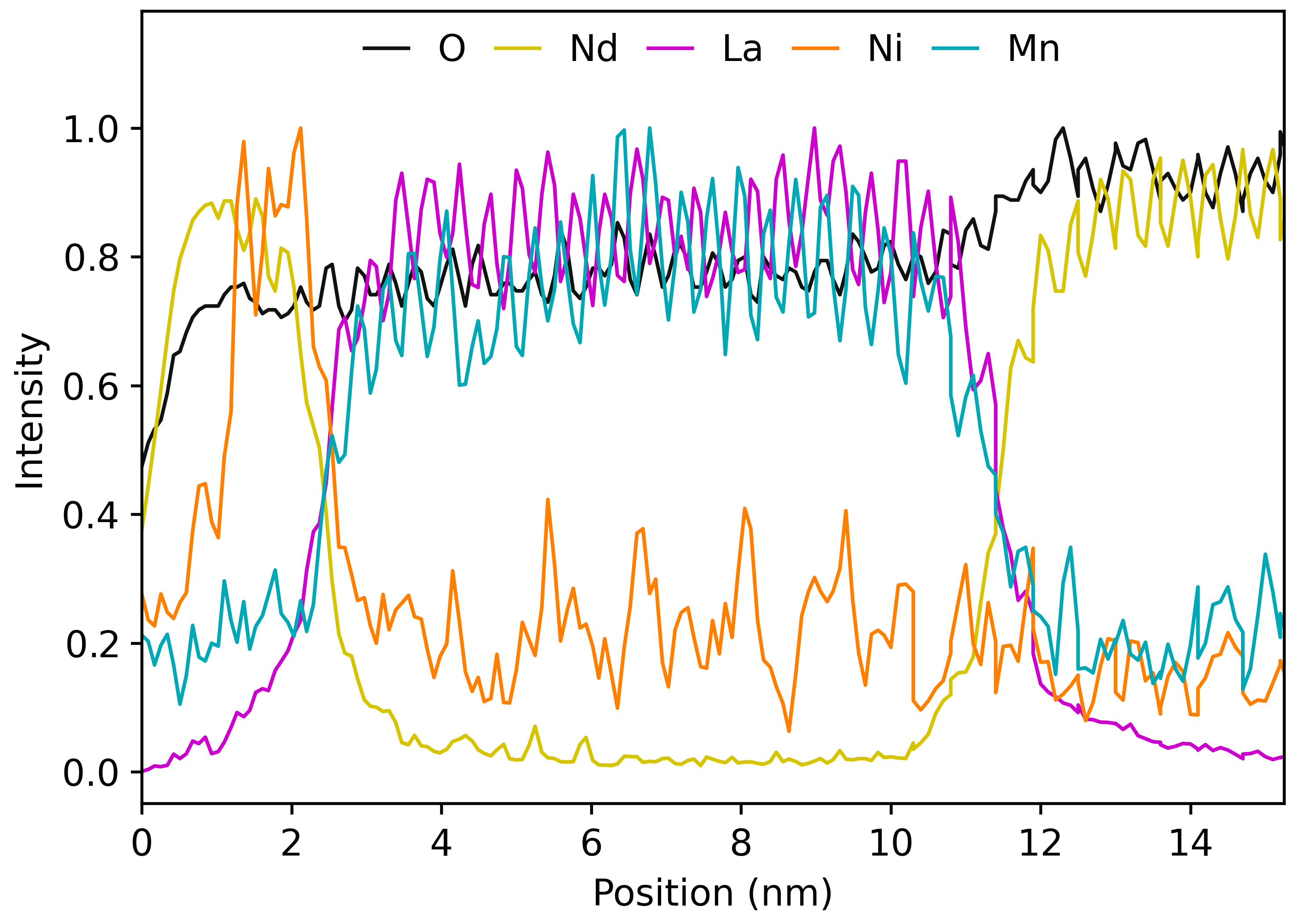}
\end{subfigure}%
\caption{(a) Profiles of the fitted Ni$^{2+,3+}$ and Mn$^{2+,3+,4+}$ oxidation-state components extracted along the vertical direction of the spectrum image shown in Fig.~\ref{fig:MapsNd}. (b) Elemental intensity profiles of O, Nd, Ni, La and Mn signals along the same direction. Each profile is normalized to its maximum.}
\label{fig:nimnprofile}
\end{figure}

The Ni-L$_{2,3}$ edge is reliably separated from the overlapping La-M$_{4,5}$ edge using explicit model-based fitting, in which the La$^{3+}$ M-edge contribution is fitted simultaneously with theoretical Ni$^{2+}$ and Ni$^{3+}$ spectral components. As shown in Fig.~\ref{fig:lanispecnno}, the resulting model reproduces the experimental spectra across different regions of the specimen defined in Fig.~\ref{fig:dfsnno}, and the localization of the recovered Ni signal to the Nd-rich region supports the robustness of the La/Ni deconvolution.  The fitted Ni$^{2+}$ and Ni$^{3+}$ components yield relative populations of $66\%$ and $34\%$ in area 1 and $61\%$ and $39\%$ in area 2, respectively, while the fitted Ni contribution in area 3 is essentially zero, as shown in Figs. \ref{fig:laniexcarea1}, \ref{fig:laniexcarea2}, \ref{fig:laniexcarea3}. The complementary La/Mn and Nd/Ni intensity profiles in Fig. \ref{fig:nimnprofile}  further indicate an interface that is abrupt within the spatial resolution of the experiment, although the Mn$^{2+}$ signal detected on the NdNiO$_3$ side suggests limited interfacial cation interdiffusion.

\begin{figure}[h]
    \centering
    \begin{subfigure}[b]{0.8\linewidth}
        \includegraphics[width=\linewidth]{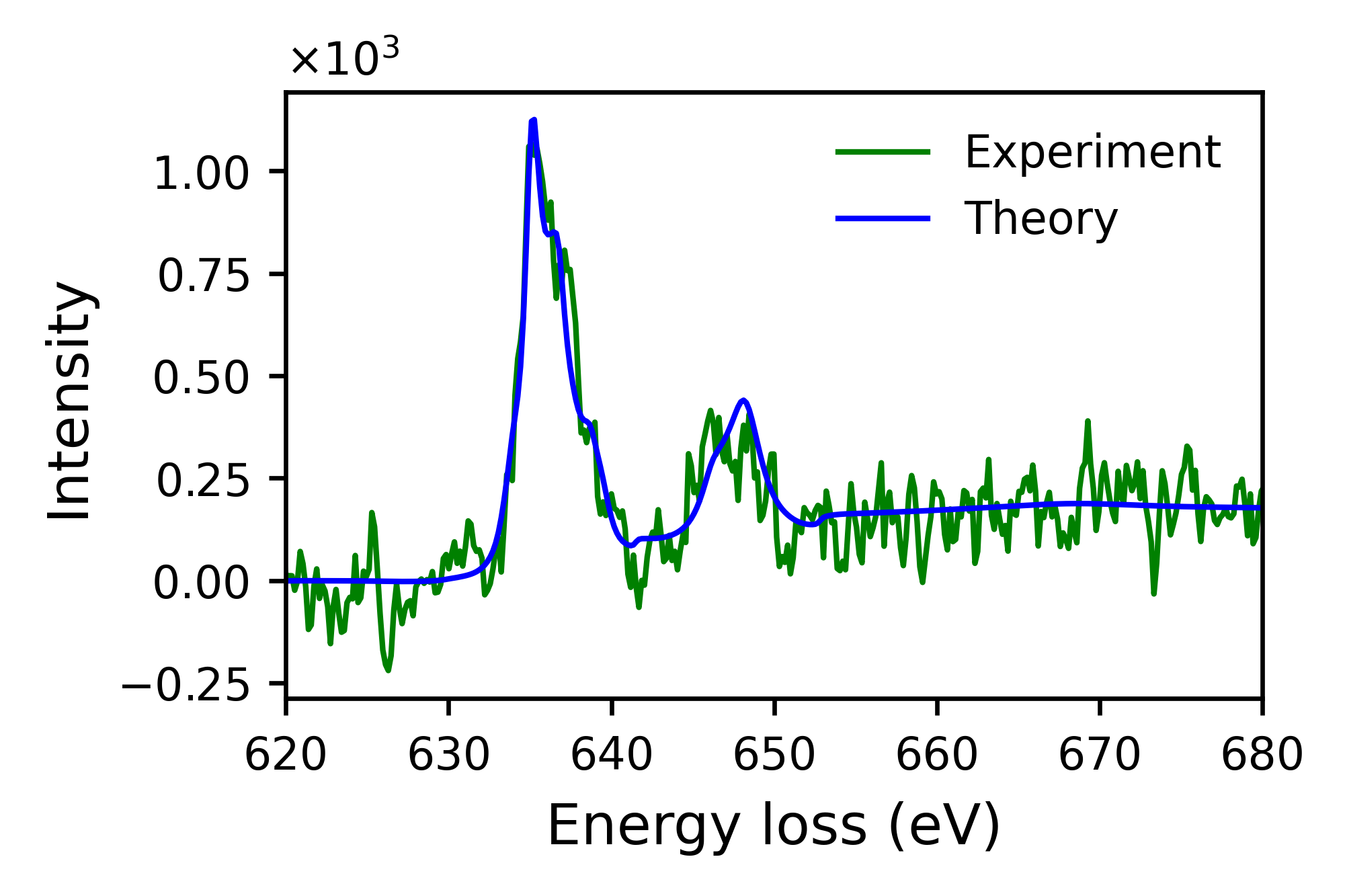}
        \caption{}
        \label{fig:nnoMnspecarea1}
    \end{subfigure}
    %\hspace{0.1em}
    \begin{subfigure}[b]{0.8\linewidth}
        \centering
        \includegraphics[width=\linewidth]{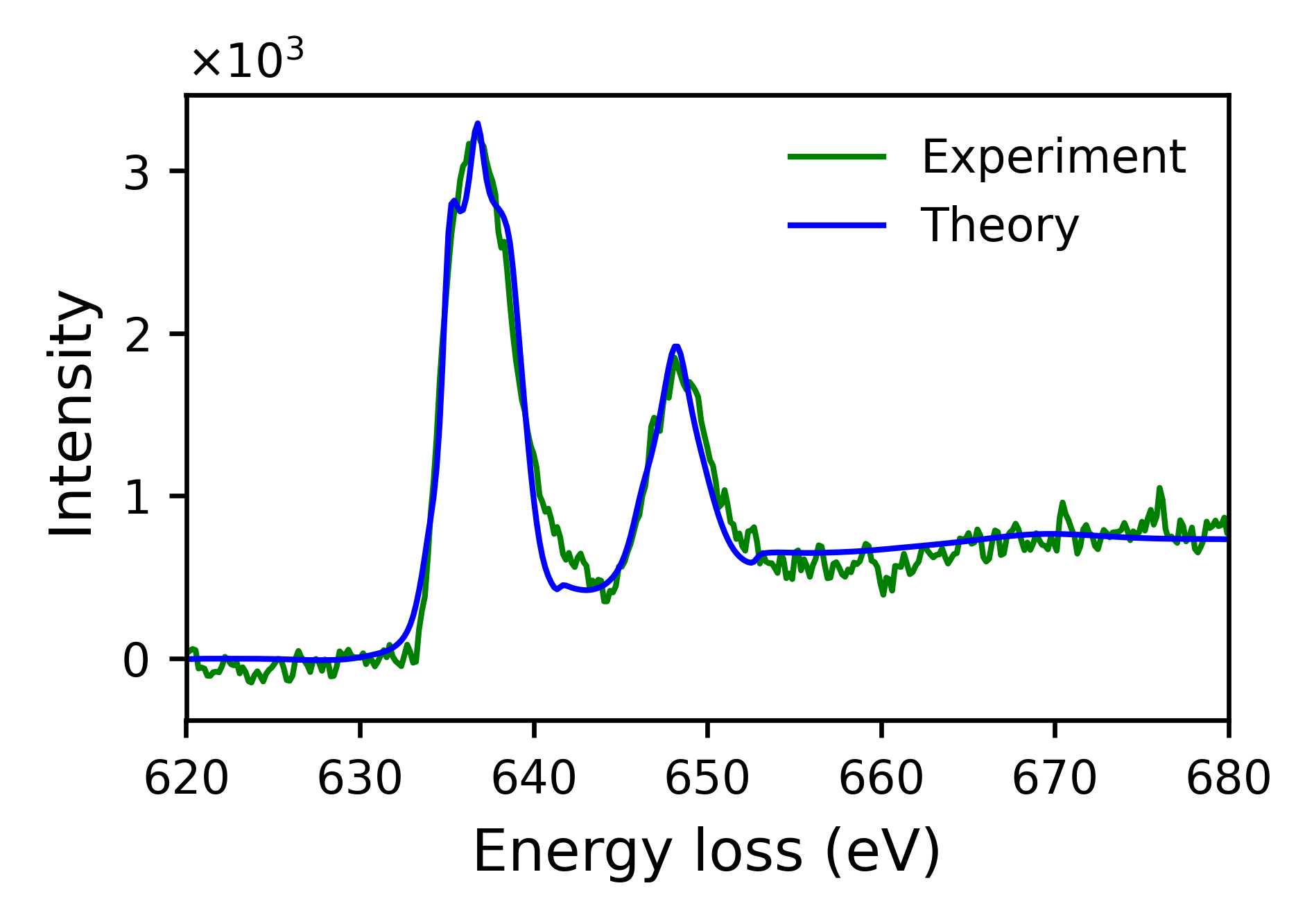}
        \caption{}
        \label{fig:nnoMnspecarea2}
    \end{subfigure}
    %\hspace{0.1em}
    \begin{subfigure}[b]{0.8\linewidth}
        \centering
        \includegraphics[width=\linewidth]{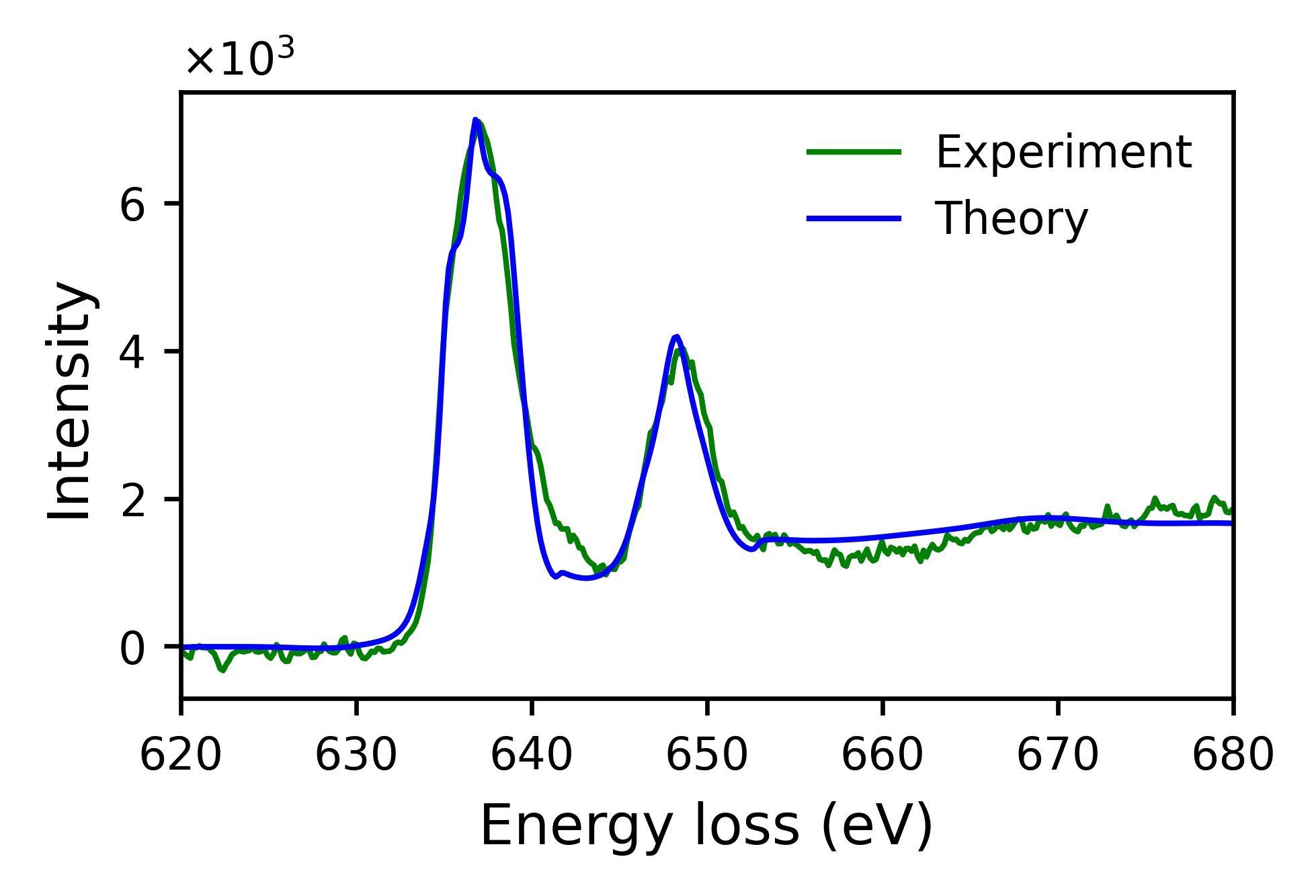}
        \caption{}
        \label{fig:nnoMnspecarea3}
    \end{subfigure}
    \caption{(a), (b), (c) are comparisons between experiment (green) and calculated models (blue) of $\mathrm{Mn}-L$ spectrum at areas 1, 2, 3, as defined in Fig.~\ref{fig:dfsnno}, respectively.}
    \label{fig:mnnno}
\end{figure}

The spectral fitting of the Mn-L$_{2,3}$ edge across the integration areas of the spectrum image, defined in Fig.~\ref{fig:dfsnno}, validates the theoretical spectral shapes (Fig.~\ref{fig:mnnno}) and also reveals a Mn$^{2+}$ contribution on the NdNiO$_3$ side. Because Mn is not a nominal constituent of NdNiO$_3$, this signal is most appropriately interpreted as evidence for limited interfacial cation interdiffusion rather than as an intrinsic nickelate valence state. From the fits in Fig.~\ref{fig:mnnno}, the relative populations of Mn$^{2+}$/Mn$^{3+}$/Mn$^{4+}$ in areas 1, 2, and 3 are $74\%/26\%/0\%$, $53\%/35\%/12\%$, and $32\%/46\%/22\%$, respectively.

\begin{figure}[!htbp]
\vspace{.5cm}
\includegraphics[height=\linewidth]{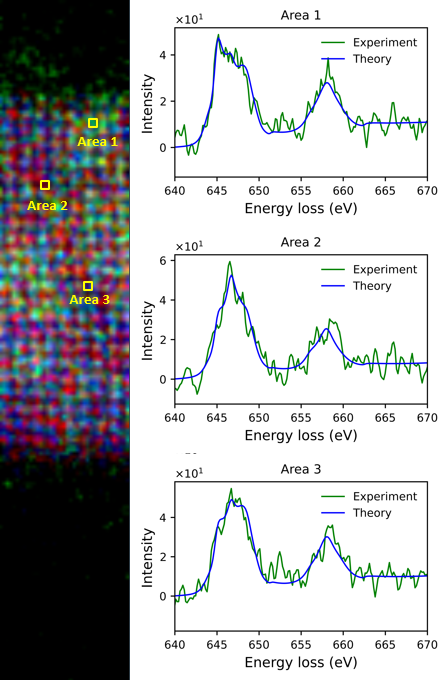}
\caption{The colormix of the Mn$^{2+, 3+, 4+}$ distribution maps is shown in green, red and blue colors, respectively. Atomic column-to-column variations in the relative contributions of the different Mn oxidation states are visible. The background-subtracted spectra from areas 1, 2 and 3 are presented together with the fitted Mn-L$_{2,3}$ inelastic differential cross section. The size of each integration area is $2\times 2$ pixels.}
\label{fig:Mncolormix}
\end{figure}

In the NdNiO$_3$/La$_{0.67}$Sr$_{0.33}$MnO$_3$/NdGaO$_3$ specimen, the Mn$^{2+}$, Mn$^{3+}$, and Mn$^{4+}$ components within the La$_{0.67}$Sr$_{0.33}$MnO$_3$ layer indicate a clear mixed-valence state, as shown in Figs.~\ref{fig:MapsNd}, \ref{fig:nimnprofile}. The Mn$^{2+}$, Mn$^{3+}$, and Mn$^{4+}$ maps are combined in a colormix representation in Fig.~\ref{fig:Mncolormix}, with the three components shown in green, red, and blue, respectively. The composite map reveals column-to-column variations in the locally dominant Mn oxidation-state contributions at near-atomic resolution. Representative Mn-$L_{2,3}$ spectra were extracted from $2\times2$ pixel regions, corresponding to approximately $1.7\times1.7$~\AA$^2$. The spectra therefore probe atomic-column-scale regions and exhibit corresponding variations in the near-edge spectral shape, which are reproduced by the fitted theoretical models. The different absolute Mn intensities between the selected regions reflect variations in the local Mn signal and are distinct from the relative Mn$^{2+}$, Mn$^{3+}$, and Mn$^{4+}$ populations obtained from the spectral decomposition.

Figure~\ref{fig:nimnprofile} combines the oxidation-state profiles of Mn and Ni with the corresponding elemental intensity profiles across the NdNiO$_3$/La$_{0.67}$Sr$_{0.33}$MnO$_3$/NdGaO$_3$ interface. On the nickelate side, the Ni signal remains predominantly Ni$^{2+}$ without further reduction towards the interface. Mn evolves toward higher average oxidation state from the NdNiO$_3$ side through the interface and into the La${_0.67}$Sr${_0.33}$MnO$_3$ layer, although the Mn signal within NdNiO$_3$ is attributed to limited cation interdiffusion. At the same time, the elemental profiles show that the O-K intensity does not vary strongly across the interface within the precision of the present experiment. A nearly constant integrated O-K intensity does not exclude changes in Ni–O hybridization or smaller oxygen-stoichiometry variations below the sensitivity of the present analysis.

Because Mn is present on the NdNiO$_3$ side through cation interdiffusion, the spatial evolution of its fitted oxidation-state distribution cannot be interpreted solely as intrinsic electronic reconstruction. The absence of a clear additional Ni reduction toward the interface likewise precludes establishing a simple one-to-one Mn oxidation/Ni reduction charge-transfer mechanism. The interface is therefore more conservatively described as involving both cation intermixing and spatially varying transition-metal valence states.

The present results show that quantitative interpretation of nickelate interfaces requires consideration of edge overlap, mixed valence, interdiffusion, and metal-oxygen hybridization. The quantitative analysis of the Ni and Mn valence states in the NdNiO$_3$ and La$_{0.67}$Sr$_{0.33}$MnO$_3$ layers, respectively, was enabled by theoretical modeling of the spectral shapes associated with the relevant oxidation states, despite the strong overlap between the La-M$_{4,5}$ and Ni-L$_{2,3}$ edges. This approach enables not only the detection of Ni, but also the quantitative determination of the relative contributions of its different valence states. The effective Ni valence below 3+ is consistent with previous reports \cite{PhysRevB.94.195127}.

%\begin{widetext}
    
%\end{widetext}

\section{Summary and outlook}

In this work, we established a quantitative framework for near-atomic-resolution oxidation-state analysis in complex oxide heterostructures by combining monochromated STEM-EELS with a theoretical calculation of the total inelastic differential cross section within the relaxed Dirac-Hartree-Fock and ligand-field multiplet frameworks \cite{deGroot2005, IatrakisBrudanin2026}. In contrast to empirical fingerprinting approaches, the present method explicitly treats the excitations to atomic-like discrete states and the continuum including many-body and crystal-field contributions. This approach enables physically grounded inference of oxidation states from transition-metal L$_{2,3}$ and rare-earth M$_{4,5}$ edges even in the presence of mixed valence and strong spectral overlap \cite{deGroot2005,PhysRevB.32.5107,PhysRevB.33.4253}.

Applied to La$_{0.67}$Sr$_{0.33}$MnO$_3$/LaCr$_{0.2}$Mn$_{0.2}$Fe$_{0.2}$Co$_{0.2}$ Ni$_{0.2}$O$_3$/SrTiO$_3$, the method resolves a pronounced redox gradient across the manganite/high-entropy-oxide stack. The analysis identifies spatially varying Mn$^{2+}$, Mn$^{3+}$, and Mn$^{4+}$ contributions together with Cr$^{3+}$, Fe$^{3+}$, Co$^{2+, 3+}$ states in the multication layer and enables La/Ni deconvolution even for a weak Ni signal, while the SrTiO$_3$ substrate remains predominantly Ti$^{4+}$. These results demonstrate selective stabilization of different transition-metal valence states and site-specific valence redistribution within the HEO layer, rather than a uniform 3+ oxidation state across all B-site cations, in agreement with the broader spectroscopy of mixed manganite/cobaltite systems \cite{Abbate1992,doi:10.1021/acs.jpcc.6b00949,Che2024}.

Applied to NdNiO$_3$/La$_{0.67}$Sr$_{0.33}$MnO$_3$/NdGaO$_3$, the framework resolves mixed Mn valence in the La$_{0.67}$Sr$_{0.33}$MnO$_3$ layer and, crucially, deconvolves the strongly overlapping La-M$_{4,5}$ and Ni-L$_{2,3}$ edges without empirical subtraction \cite{PhysRevB.90.195140}. The recovered Ni signal remains localized in the Nd-rich region, supporting the robustness of the La/Ni separation and enabling interpretation of the intrinsic Ni fine structure and local electronic structure.

In NdNiO$_3$/La$_{0.67}$Sr$_{0.33}$MnO$_3$/NdGaO$_3$, an effective Ni valence below $3+$ is consistent with the correlated description of NdNiO$_3$ \cite{PhysRevB.94.195127}. The fitted Ni$^{2+}$ and Ni$^{3+}$ populations show a modest spatial variation across the interface, while the average Ni valence remains approximately constant within the precision of the present analysis. The Mn oxidation-state distribution varies strongly across the same region, but the detected Mn signal within the NdNiO$_3$ side indicates cation interdiffusion and therefore complicates a direct interpretation in terms of electronic charge transfer. The present results are interpreted as a combination of cation intermixing and spatially varying local electronic structure rather than as a simple one-to-one Mn oxidation/Ni reduction mechanism.

A future direction is the application of the same workflow to other multication and rare-earth oxides, including Ce-containing systems, as well as Fe-, Co-, and Ti-based oxide interfaces where local redox gradients and hybridization effects are central to functionality \cite{PhysRevB.32.5107,FILATOVA201727}. Coupling near-atomic-resolution EELS with operando or bias-dependent experiments would provide a route to directly monitor how oxidation state, covalency, and defect populations evolve under external driving forces, which is likely to be essential for understanding nickelates and catalytic perovskites under working conditions \cite{doi:10.1021/acs.jpcc.6b00949, baeumer2017anomalous,D3CC00325F,Baeumer2021}.

\begin{acknowledgments}
We would like to thank N. Tasios, P. Longo, V. Brudanin and J. Smeets for useful discussions.
\end{acknowledgments}

\bibliography{EELSMultipletTheory}% Produces the bibliography via BibTeX.

\end{document}